\documentclass[12pt]{article}

\usepackage{amsmath,amsfonts}
\usepackage{hyperref}
\usepackage[dvips]{color}
\usepackage{graphicx}
\makeatletter\@addtoreset{equation}{section}\makeatother

\def\bC {\mathbb{C}}

\def\bP {\mathbb{P}}
\def\bR {\mathbb{R}}
\def\bZ {\mathbb{Z}}

\def\CP{\bC\bP}

\newcommand{\be}{\begin{equation}}
\newcommand{\ee}{\end{equation}}
\newcommand{\bea}{\begin{eqnarray}}
\newcommand{\eea}{\end{eqnarray}}

\newcommand{\vev}[1]{{\left< {#1} \right>}}

\newcommand{\eqn}[1]{(\ref{#1})}

\def\bsp{\be\begin{split}}

\def\G{\Gamma}
\def\D{\Delta}
\def\a{\alpha}
\def\b{\beta}

\def\k{\kappa}
\def\d{\delta}

\def\m{\mu}
\def\n{\nu}
\def\r{\rho}
\def\l{\lambda}

\def\p{\partial}

\def\bR {\mathbb{R}}
\def\bZ {\mathbb{Z}}

\newcommand{\Tr}{{\rm Tr\,}}

\newcommand{\cL}{{\mathcal L}}
\newcommand{\cM}{{\mathcal M}}
\newcommand{\cN}{{\mathcal N}}

\newcommand{\cO}{{\mathcal O}}
\newcommand{\cS}{{\mathcal S}}

\renewcommand{\title}[1]{\vbox{\center\LARGE{#1}}\vspace{5mm}}
\renewcommand{\author}[1]{\vbox{\center#1}\vspace{5mm}}
\newcommand{\address}[1]{\vbox{\center\em#1}}
\newcommand{\email}[1]{\vbox{\center\tt#1}\vspace{10mm}}

\begin{document}
\bibliographystyle{utphys}  
\begin{titlepage}
\begin{center}
\hfill {\tt HU-EP-08/43}\\
\vspace{20mm}
\title{\bf Vortex Loop Operators,\\
M2-branes and Holography} 

\author{\large Nadav Drukker$^{1,a}$, 
Jaume Gomis$^{2,b}$
and 
Donovan Young$^{1,c}$}

\address{$^1$Humboldt-Universit\"at zu Berlin, Institut f\"ur Physik\\
Newtonstra{\ss}e 15, D-12489 Berlin, Germany
\\\medskip
$^2$Perimeter Institute for Theoretical Physics\\
Waterloo, Ontario N2L 2Y5, Canada}

\email{$^a$drukker, $^c$dyoung @physik.hu-berlin.de\\
$^b$jgomis@perimeterinstitute.ca}

\end{center}
\abstract{
\medskip\medskip
\noindent
{\normalsize
We construct   vortex loop operators in the three-dimensional $\cN=6$ supersymmetric 
Chern-Simons theory recently constructed by Aharony, Bergman, Jafferis 
and Maldacena.  These  disorder loop operators are specified by a  vortex-like
singularity  for the scalar and gauge fields along  a one dimensional curve in spacetime.
We identify the $1/2$, $1/3$ and $1/6$ BPS loop operators in the Chern-Simons  
theory with excitations of M-theory corresponding to M2-branes ending along a 
curve on the boundary of $AdS_4\times S^7/\bZ_k$. The vortex loop operators 
can also be given   a  purely geometric description  in terms of     regular  
``bubbling'' solutions of eleven dimensional  supergravity which are asymptotically $AdS_4\times S^7/\bZ_k$.}
}

\end{titlepage}

\tableofcontents
\addtolength{\parskip}{1ex}

\vfill\eject

\section{Introduction} 

Three dimensional Chern-Simons theory is a topological field theory whose
only known observables are Wilson loop operators, which are  supported on knots and links in the three manifold.
Chern-Simons theory coupled to matter --- which describes a wealth of physical phenomena ---
has a much richer set of observables, that can be used to characterize the physical properties of the system.

In this paper we construct a novel class of operators in Chern-Simons theories 
coupled to matter. We do this in the ${\cal N}=6$ supersymmetric Chern-Simons 
theory of Aharony, Bergman, Jafferis and Maldacena \cite{Aharony:2008ug}, 
yet our construction generalizes to any Chern-Simons theory coupled to matter 
fields, and may find interesting applications elsewhere, and serve as order 
parameters for new phases in three dimensional theories.

The operators we construct --- which we will denote by $V_C$ --- are supported 
on a curve $C$ in the three dimensional manifold in which the Chern-Simons-matter 
theory is defined, and are therefore loop operators. Unlike the more familiar Wilson 
loop operators, $V_C$  are disorder loop operators, defined by a path integral 
with certain singularities for the fields of the theory along the loop $C$. 

These operators are characterized by a vortex-like singularity for the 
Chern-Simons-matter fields near the location of the loop $C$.  Since a 
vortex in a Chern-Simons-matter theory describes a particle with arbitrary 
statistics, the insertion of a  loop operator  $V_C$ has the effect of creating 
a probe  anyon with a  worldline  specified by the curve $C$, with which the 
theory is probed. They can also be viewed as singular limits of solitonic 
vortex solutions that exist in some Chern-Simons theories coupled to matter 
\cite{Hong:1990yh,Jackiw:1990aw,Jackiw:1990pr}.

We present a family of loop operators $V_C$ in the  $U(N)_k\times U(N)_{-k}$ 
${\cal N}=6$ Chern-Simons theory of \cite{Aharony:2008ug} which preserve 
$1/2$, $1/3$ or $1/6$ of the twenty-four  supercharges of the vacuum. All 
these operators will have singularities for some of the gauge fields and 
some of the scalar fields along the curve $C$.
These operators are labeled by certain parameters which specify the 
possible supersymmetric,  codimension two singularities allowed in the theory. 
This data is rather rich, giving a high dimensional moduli space. The one-half 
BPS codimension two singularities we find are reminiscent of the ones corresponding 
to disorder surface operators in ${\cal N}=4$ SYM \cite{Gukov:2006jk} (see also \cite{Gomis:2007fi}), 
 whose data parametrizes the moduli space of solutions of the Hitchin equations 
 in the presence of   codimension two singularities.

In the second part of the paper, we provide the explicit bulk description of these 
novel loop operators in ${\cal N}=6$ Chern-Simons theory by identifying them 
with  excitations of M-theory in $AdS_4\times S^7/\bZ_k$, providing strong 
evidence for the proposal in \cite{Aharony:2008ug}   that ${\cal N}=6$ 
Chern-Simons theory is the holographic description of M-theory with 
$AdS_4\times S^7/\bZ_k$ boundary conditions.

We identify the loop operators $V_C$ in ${\cal N}=6$ Chern-Simons theory with configurations of M2-branes in $AdS_4\times S^7/\bZ_k$ ending on the boundary of $AdS_4$ along a curve   $C$, the singular locus of the  
 loop operators.  For all these solutions we find an explicit map  between the data characterizing the loop operators   in the gauge theory and the data characterizing the M2-brane configuration  in $AdS_4\times S^7/\bZ_k$. We further show  that a class  of  asymptotically $AdS_4\times S^7$ solutions  constructed by Lunin \cite{Lunin:2007ab} can be appropriately orbifolded to yield the backreacted description of our  M2-brane configurations. These non-singular asymptotically $AdS_4\times S^7/\bZ_k$ ``bubbling" solutions of eleven dimensional supergravity provide  the purely gravitational description of our 1/2 loop operators $V_C$.

At weak 't Hooft coupling  we compute --- in the semiclassical approximation --- the expectation value of a loop operator $V_C$, the correlator of $V_C$ with a chiral primary operator as well as as the correlator of $V_C$ with the stress tensor of ${\cal N}=6$ Chern-Simons theory. Using the M2-brane description of loop operators, we compute using bulk supergravity methods  the loop operator expectation value and  the  correlator of a loop operator with a chiral primary operator in the strong coupling regime. The remarkable agreement found in the case of $\cN=4$ SYM 
between the semiclassical gauge theory computation and the bulk strong coupling computation for the corresponding correlators of  surface operators \cite{Drukker:2008wr} does not hold in this case.  

The loop operators constructed in this paper together with the Wilson loop operators constructed in \cite{Drukker:2008zx,Chen:2008bp,Rey:2008bh} (and foretold already in \cite{Gaiotto:2007qi}) provide a rich set of non-local observables in ${\cal N}=6$ Chern-Simons theory, which can be used to study the phase structure of these Chern-Simons-matter theories.

The plan of the rest of the paper is as follows. In section \ref{sec-CS} we classify and explicitly construct 1/2, 1/3 and 
 1/6 BPS loop operators in ${\cal N}=6$ Chern-Simons theory with Abelian and non-Abelian gauge groups.
 These operators are constructed in terms of codimension two singularities of the theory on ${\bR^3}$ as well as vacua of the theory on $AdS_2\times S^1$. We then calculate in the leading semiclassical approximation the expectation value of $V_C$ and the correlator of $V_C$ with a chiral primary operator and the stress tensor.
 Section  3 contains the bulk gravitational description of the loop operators studied in section \ref{sec-CS}. We identify the M2-brane configuration in $AdS_4\times S^7/\bZ_k$ corresponding to $V_C$ as well as the ``bubbling" supergravity solution description of $V_C$. We also calculate using our probe M2-brane description the expectation value of $V_C$ as well as the correlator of $V_C$ with a chiral primary operator. A discussion  and summary of our results can be found in section $4$. Some technical details and computations are  relegated to appendices.

\section{Vortex Loop Operators in $\cN=6$ Chern-Simons Theory}
\label{sec-CS}
 
In this section we construct supersymmetric disorder loop operators  
in $\cN=6$ supersymmetric Chern-Simons theory. These operators are 
supported on a  curve $C$ in spacetime, and   will be denoted  by  $V_C$. 
Physically, a  disorder loop operator $V_C$  inserts into the system an 
external particle, with which the theory can be 
probed. As we shall see, the field configuration near  $V_C$ is that 
of a vortex, and since a particle described by a vortex in Chern-Simons theory coupled to matter can acquire any statistics, the particle inserted by  $V_C$ is an anyon.
 
The disorder loop operator $V_C$ in a three dimensional field  
theory in $\bR^{3}$ is constructed  by specifying a singularity for 
the fields in the theory near the curve $C$ in spacetime.    
The only restriction is that the singular field configuration solves the 
equations of motion of the theory in $\bR^{1,2}\backslash C$. The problem 
of constructing disorder loop operators gets mapped to the problem of 
classifying the codimension two singularities  for the  fields in the theory%
\footnote{We write $\bR^3$ even-though the calculation is this section (apart for Subsections~\ref{sec-circle} and~\ref{sec-CS-AdS2}) is done in Lorentzian signature. The M-theory dual in Section~\ref{sec-M} is described with Euclidean signature, apart for the supersymmetry calculation in Appendix~\ref{app-M2-SUSY}.}
 in $\bR^{3}$.

$\cN=6$ supersymmetric Chern-Simons theory has  $U(N)\times U(N)$ gauge symmetry%
\footnote{For gauge group $SU(2)\times SU(2)$ it is equivalent to Bagger-Lambert-Gustavsson theory \cite{Gustavsson:2007vu,Bagger:2007jr} where vortex solutions were also recently found \cite{KIm:2008cp}.}
 and the bosonic fields are a pair of gauge fields $A$ and $\hat{A}$ and four complex scalar fields $C^I=(C^1,C^2,C^3,C^4)$ 
transforming in the bifundamental representation of the gauge 
group. The  Lagrangian for these fields is given by\footnote{We have rescaled the matter fields such that $k$ appears as an overall factor in the Lagrangian.} 
\be
\begin{aligned}
{{\cal L}}= \frac{k}{4 \pi} \varepsilon^{\mu\nu\lambda} \mathrm{Tr} &\left(
 A_{\mu} \partial _{\nu} A_\lambda + \frac{2i}{3} A_{\mu} A_{\nu} A_{\lambda}
- \hat{A}_{\mu} \partial_{\nu} \hat{A}_{\lambda}
- \frac{2i}{3} \hat{A}_{\mu} \hat{A}_{\nu} \hat{A}_{\lambda} \right)\cr
-& k\,\mathrm{Tr}\, D_{\mu} C_I^{\dagger} D^{\mu} C^I -V_{pot}\,, 
  \label{ABJM}
\end{aligned}
\ee
where 
\be
D_\mu C^I= \partial_\mu C^I-iA _\mu\, C^I+iC^I\hat{A} _\mu\,
\ee
and $V_{pot}$ denotes a sextic scalar potential, whose explicit form can be found in 
\cite{Aharony:2008ug,Benna:2008zy}. The theory depends on the integer $k$, which determines the level of the Chern-Simons interactions. For $k\gg1$, the theory has a weakly coupled expansion controlled by $1/k$.
One can further define an 't Hooft limit, where $N\rightarrow \infty, k\rightarrow \infty$ with $\lambda=N/k$ kept fixed.

The equations of motion for the gauge fields with bosonic sources are
\be
\begin{aligned}
\label{eomnonab}
\frac{1}{4\pi}\varepsilon^{\mu\nu\lambda}F_{\mu\nu}&=iD^\lambda C^I C_I^\dagger -iC^ID^\lambda C_I^\dagger 
\\
\frac{1}{4\pi}\varepsilon^{\mu\nu\lambda}\hat{F}_{\mu\nu}&=iC_I^\dagger D^\lambda C^I-iD^\lambda C_I^\dagger C^I\,,
\end{aligned}
\ee
where
\be \label{fnonab}
F_{\mu\nu}=\partial_\mu A_\nu-\partial_\nu A_\mu+i[A_\mu,A_\nu]\qquad
\hat{F}_{\mu\nu}=\partial_\mu \hat{A}_\nu-\partial_\nu \hat{A}_\mu+i[\hat{A}_\mu,\hat{A}_\nu]\,.
\ee
Disorder loop operators in this theory are characterized by the 
allowed codimension two singularities for  $A$,  $\hat{A}$  and  $C^I$.

 In this paper we are interested in supersymmetric loop 
operators, which greatly simplifies the analysis. The Chern-Simons theory in 
\cite{Aharony:2008ug} is invariant under ${\cal N}=6$ Poincar\'e supersymmetries, which 
we parametrize by three dimensional spinors $\epsilon_{IJ}=-\epsilon_{JI}$, where $I,J=1,\cdots,4$. A disorder loop operator is supersymmetric when the supersymmetry variation of all the fields vanishes in the background it creates. The supersymmetry variation of the bosonic fields is automatically zero, so we need to examine the supersymmetry variation of the fermions, which is given by  \cite{Gaiotto:2008cg,Terashima:2008sy, Bandres:2008ry}
\be
  \delta \psi_I = - \gamma^\mu \epsilon_{IJ} D_\mu C^J
+{2 \pi} \left(
-\epsilon_{IJ} (C^K C_K^{\dagger} C^J-C^J C_K^{\dagger} C^K)
+2 \epsilon_{KL} C^K C_I^\dagger C^L
\right)\,.
\label{poincare}
\ee
These equations must be supplemented with the equations of motion for the gauge fields (\ref{eomnonab}).

The theory in \cite{Aharony:2008ug} is also invariant under ${\cal N}=6$ conformal supersymmetries, which are parametrized by three dimensional spinors $\eta_{IJ}=-\eta_{JI}$, where $I,J=1,\cdots,4$. A loop operator  invariant under conformal supersymmetries is described by a bosonic field configuration with vanishing \cite{Bandres:2008ry} 
\begin{eqnarray}
\delta \psi_I &=& - \gamma^\mu \gamma^\nu x_\nu\eta_{IJ} D_\mu C^J
+{2 \pi}\gamma^\nu x_\nu \left(
-\eta_{IJ} (C^K C_K^{\dagger} C^J-C^J C_K^{\dagger} C^K)
+2 \eta_{KL} C^K C_I^\dagger C^L\right)
\cr&&
-\eta_{IJ}C^J\,.
\label{conformal}
\end{eqnarray}
Altogether, the  ${\cal N}=6$ Chern-Simons theory in  \cite{Aharony:2008ug} is invariant under the $OSp(6|4)$ supergroup. We will now construct families of  supersymmetric loop operators that are invariant under various subgroups of $OSp(6|4)$.

\subsection{Loop Operators in the $U(1)\times U(1)$ Theory}
\label{secabelia}

We start by describing the operator  $V_C$ corresponding to 
inserting a static particle in the theory with $U(1)\times U(1)$ gauge group. 
For a static particle the curve $C$ is  a straight line $C=\bR\subset \bR^3$. 
We choose coordinates $(t,z,\bar{z})$ such that the line 
is defined by $z=0$ and parametrized by $t$. The straight line --- together with the circle ---
are the two maximally symmetric curves in $ \bR^3$. They are both invariant under 
an $SU(1,1)\times U(1)_l$ subgroup of the three dimensional 
conformal group $SO(2,3)$. 

Once the singularity for the straight line is understood, one can then
construct the loop operator $V_C$  for an arbitrary curve  $C\subset \bR^3$, by 
treating $(z,\bar{z})$ as local coordinates in the  normal bundle of $C$. For a curve $C$ other than $\bR$ or $S^1$, the 
$SU(1,1)\times U(1)_l$ symmetry is broken.

\subsubsection*{$\bullet\quad ${$1/2$ BPS Loop Operators}}

A maximally supersymmetric loop operator in ${\cal N}=6$ Chern-Simons theory is obtained by allowing a single complex scalar field to acquire a singularity near the curve $C$. Exciting multiple scalar fields preserves less supersymmetry.\footnote{Unless all scalar fields are proportional to each other, in which case they   preserve the same supersymmetry as the case of a single scalar.}  
Therefore, we first consider the following codimension two scalar field singularity%
\footnote{We focus on static configurations in this paper and do not consider any possible time dependence for the fields.}
\be
C^1= f(z,\bar{z})\,,
\label{singus}
\ee
  $f(z,\bar{z})$ is an arbitrary  function that develops a singularity at $z=0$, the location of the operator $V_C$. 
The choice of a complex scalar field breaks the $SU(4)$ R-symmetry of the theory down to $SU(3)\times U(1)_R$. 

The operator $V_C$ is supersymmetric if the field configuration produced by $V_C$ gives a vanishing supersymmetry variation for the Fermi fields (\ref{poincare}). 
It is convenient to decompose the supersymmetries according to their helicity in the $z$-plane, so that  $\epsilon_{IJ}=\epsilon^+_{IJ}+\epsilon^-_{IJ}$,   where the helicity components satisfy
\be
\gamma^z\epsilon^+_{IJ}=0\qquad\qquad 
\gamma^{\bar z}\epsilon^-_{IJ}= 0\,.
\label{heli}
\ee
Moreover, the spinors satisfy a reality condition, where complex conjugation raises their indices. In our basis it also flips their helicity
\be
(\epsilon_{IJ}^\mp)^*=\epsilon^{IJ}{}^\pm={1\over 2}\epsilon^{IJKL}\epsilon_{KL}^\pm\,.
\label{complex}
\ee

In the Abelian theory 
only the first term in (\ref{poincare}) is non-vanishing. 
Imposing that  $V_C$ leaves invariant the three supercharges parametrized by
$\epsilon^+_{1I}$
gives rise to the following  BPS equations
\be
D_{\bar z}C^1=0\qquad \qquad D_{t}C^1=0 \,.
\label{BPS}
\ee
The BPS equations restrict the other three scalar fields $C^2$, $C^3$ and $C^4$ to be (covariantly) constant.

Due to  equation (\ref{complex}), any solution of these BPS equations is automatically also  invariant under three more supersymmetry variations with parameters 
$\epsilon^{1JKL}\epsilon^-_{KL}$, yielding a configuration 
 invariant under six real Poincar\'e supercharges. Therefore, solutions to (\ref{BPS}) preserve one-half of the Poincar\'e supersymmetries.
Explicitly, they are invariant under the supersymmetry transformations labeled by
\be
\left\{\epsilon^+_{12}\,,
\epsilon^+_{13}\,,
\epsilon^+_{14}\,,
\epsilon^-_{23}\,,
\epsilon^-_{24}\,,
\epsilon^-_{34}\right\}.
\label{six}
\ee

The BPS equations (\ref{BPS}) must be supplemented with the equations of motion 
for the gauge fields. 
In the Abelian theory, the  matter fields couple only to a linear combination of the gauge fields through 
\be
D_\mu C^I= \partial_\mu C^I-iA^-_\mu\, C^I\,,
\ee
where 
\be
A^+=A+\hat{A}\qquad A^-=A-\hat{A}\,.
\ee
The other gauge field, $A^+$, appears in the action only in a  Chern-Simons term. 
The equations of motion for the gauge fields (\ref{eomnonab}) are now
\be
\begin{aligned}
\label{eompm}
\frac{1}{8\pi}\varepsilon^{\mu\nu\lambda}F_{\mu\nu}^+&=iD^\lambda C^I C_I^\dagger-iC^ID^\lambda C_I^\dagger\\
\varepsilon^{\mu\nu\lambda}F_{\mu\nu}^-&=0\,.
\end{aligned}
\ee

The   static solutions  of the BPS   equations (\ref{BPS})  are   given by
\be
C^1=f(z)\qquad\qquad A^-=0\,,
\label{singusc}
\ee
where $f(z)$ is an arbitrary holomorphic function that develops a singularity at $z=0$. 
This scalar field singularity (\ref{singusc}) together with the equation of motion for $A^+$   (\ref{eompm}) requires that we turn on an electric field
\be
F_{tz}^+={4\pi}f'(z)\bar{f}(\bar{z})\,,
\ee
so we may take
\be
A^+_t=-{4\pi}|f|^2\,.
\ee
Note though, that the equations of motion do not restrict the holomorphic component of the $A^+$ gauge field, allowing it to take the general form
\be
A^+_z=g(z)\,,
\label{singusb}
\ee
where  $g(z)$ is an arbitrary holomorphic function that develops a singularity at $z=0$ 
and the antiholomorphic component is its complex conjugate 
$A^+_{\bar z}=\bar A^+_z$. 
Therefore, the most general loop operator $V_C$ preserving one-half of the twelve Poincar\'e supersymmetries  in ${\cal N}=6$ Chern-Simons  theory is labeled by a pair of holomorphic  functions --- $f(z)$ and $g(z)$ ---  which are  singular at $z=0$.   

The straight line is invariant under scale transformations, which raises the possibility that the disorder operator $V_C$ be also scale invariant. Using the fact that the scalar field and  gauge field have scaling dimension $1/2$ and $1$ respectively, requiring conformal invariance fixes the strength of the singularity   characterizing $V_C$ (\ref{singusc}), (\ref{singusb}) to be%
\footnote{It also imposes $C^2=C^3=C^4=0$.}
\be
C^1={\beta\over \sqrt{z}}\, \qquad \qquad\qquad A^+_z=-i{\alpha \over 2kz}\,. 
\label{singuconf}
\ee
Therefore, this operator is labeled by two parameters $(\alpha,\beta)$. $\beta$ is a positive real number, as the phase of $C^1$ can be eliminated by a $U(1)$ gauge transformation. Likewise, the imaginary part of $\alpha$, which corresponds to a radial gauge field, can be removed, so $\alpha$ is also real, and gives the holonomy around the vortex. Since the theory is invariant under large gauge transformations, $\alpha$ is an angular variable. The allowed large gauge transformations depend on the level $k$, so that with the factor of $2k$ in the denominator of \eqn{singuconf}, $\alpha$ has  unit period \cite{Aharony:2008ug}.

We note from (\ref{singuconf}) that  the scalar field $C^1$ is not single-valued, as it changes sign upon encircling $V_C$.  Such discontinuities may seem puzzling at first, but they are rather ubiquitous in theories  with disorder operators, such as the discontinuity induced  on a   scalar field  by a $\bZ_2$ twist field in two dimensional conformal field theory. 
This discontinuity is consistent as long as the correlation functions of physical operators are well defined. As we shall explain more fully in Section~\ref{sec-local-CS}, this discontinuity does not lead to any pathologies for even $k$. The situation for odd $k$ is more complicated, as in this case there are gauge invariant operators in ${\cal N}=6$ Chern-Simons theory that are not single valued when encircling $V_C$, 
which would lead one to conclude that the vortex loop operators are unphysical for odd $k$. 
As we explain in Section~\ref{sec-nonabelian}, in the non-Abelian theory it is possible to 
have vortices also for odd $k$.

With the specific form of the singularity (\ref{singuconf}), the bosonic symmetry preserved by $V_C$ is $SU(1,1)\times U(1)_d\times SU(3)$, where $U(1)_d$ is a diagonal combination of a space-time and $R$-symmetry.\footnote{It is the diagonal sum of $U(1)_l \subset SO(2,3)$ and $U(1)_R\subset SU(4)$.} 
Furthermore, invariance under supersymmetry and conformal symmetry implies that the operators $V_C$ preserve one-half of the twelve conformal supersymmetries of the theory. Therefore the singularity (\ref{singuconf}) is invariant under the six superconformal transformations with parameters
 \be
\left\{\eta^+_{12}\,,
\eta^+_{13}\,,
\eta^+_{14}\,,
\eta^-_{23}\,,
\eta^-_{24}\,,
\eta^-_{34}\right\}.
\label{sixc}
\ee
This is   verified directly in Appendix~\ref{app-Ss} using the conformal supersymmetry transformations \eqn{conformal}.

Thus, we have constructed ${1/2}$ BPS loop operators $V_C$ for the theory with gauge group   $U(1)\times U(1)$. They are described by the singularity (\ref{singuconf}), and are invariant under an $SU(1,1|3)$ subgroup of the $OSp(6|4)$ symmetry of the theory. 
The ${1/2}$ BPS loop operators  $V_C$ are labeled by two real parameters $(\alpha,\beta)$.

\subsubsection*{$\bullet\quad  ${$1/3$ BPS Loop Operators}}

Other interesting operators $V_C$ preserving less than one-half of the Poincar\'e supersymmetries can be constructed by exciting more than a single scalar field. 

Imposing that  the operator $V_C$ leaves invariant the two supersymmetry transfomations 
with parameters $\epsilon^+_{13}$ and $\epsilon^+_{14}$
gives rise to the following  BPS equations\footnote{The supersymmetry conditions allow the scalars $C^3$ and $C^4$ to be arbitrary constants, but we will set them to zero, which is also the only conformally invariant constant.}
\be
D_{\bar z}C^1=D_{z}C^2=0\qquad \qquad D_{t}C^1=D_{t}C^2=0 \,.
\label{BPSa}
\ee
Due to  equation (\ref{complex}), any solution of the BPS equations is automatically also  invariant under two more supersymmetry transformations labeled by 
$\epsilon^-_{23}$ and $\epsilon^-_{24}$, yielding a configuration 
 invariant under four real Poincar\'e supercharges. Therefore, solutions to (\ref{BPSa}) preserve one-third  of the Poincar\'e supersymmetries. 
Explicitly, they are parametrized by
\be
\left\{\epsilon^+_{13}\,,
\epsilon^+_{14}\,,
\epsilon^-_{23}\,,
\epsilon^-_{24}\right\}.
\label{four}
\ee
As before, we should also solve the equations of motion for the gauge fields (\ref{eomnonab}).

The   static solutions  of the BPS   equations (\ref{BPSa})  are   given by
\be
C^1=f_1(z)\qquad\qquad C^2=f_2(\bar{z})\qquad\qquad A^+_z=g(z)\qquad\qquad A^-=0\,,
\ee
where $f_1(z)$ and $g(z)$ are arbitrary holomorphic functions and $f_2(\bar{z})$ is an antiholomorphic function all of which have singularities at $z=0$. An electric field for $A^+$ must also be turned on, which can be represented by the gauge potential
\be
A_t^+=-{4\pi}\left(|f_1|^2-|f_2|^2\right)\,.
\ee

If we further demand that the singularity produced by $V_C$ is scale invariant, then the form of the singularity is fixed to be
\be
C^1={\beta_1\over \sqrt{z}}\qquad\qquad C^2={\beta_2\over \sqrt{{\bar z}}}\qquad\qquad 
A^+_z=-i{\alpha\over 2kz}\,.
\label{singuthird}
\ee 
Only the relative phase of the two complex parameters $\beta_1$ and $\beta_2$ is physical,  as a  $U(1)$ gauge transformation leads to the identification $(\beta_1,\beta_2)\simeq e^{i\theta}(\beta_1,\beta_2)$. 
Therefore, these operators are labeled by $(\alpha,\beta_1,\beta_2)/U(1)$, where the $U(1)$ acts by shifting the phase of $\beta_1$, $\beta_2$ and leaves $\alpha$ invariant.

The bosonic symmetry preserved by these operators is $SU(1,1)\times SU(2)\times U(1)_{d'}$, where $U(1)_{d'}$ is a diagonal combination  of a  space-time and $R$-symmetry.\footnote{It is the diagonal sum of $U(1)_l \subset SO(2,3)$ and a $U(1)_{R'}\subset SU(4)$ under which $C^1$ and $C^2$  have charges $(+1,-1)$ respectively.} The singularity (\ref{singuthird})
preserves one-third of the conformal supersymmetries. From equation (\ref{conformal}) it follows that the singularity (\ref{singuthird}) is invariant under the four conformal supercharges labeled by
 \be
\left\{\eta^+_{13}\,,
\eta^+_{14}\,,
\eta^-_{23}\,,
\eta^-_{24}\right\}\,.
\label{fourc}
\ee
Thus, we have constructed ${1/3}$ BPS loop operators $V_C$, described by the singularity (\ref{singuthird}), which are    invariant under an $SU(1,1|2)$ subgroup of the $OSp(6|4)$ symmetry of the theory. 
The ${1/3}$ BPS loop operators  $V_C$ when the gauge group is $U(1)\times U(1)$ are labeled by $(\alpha,\beta_1,\beta_2)/U(1)$.

Our discussion throughout this paper is for the theory with general $k$, but we would like to point out that for $k=1,2$ the theory is expected to have enhanced supersymmetry --- $\cN=8$ ---  with a total of thirty-two real supercharges instead of twenty-four \cite{Aharony:2008ug}. The $1/2$ BPS vortex loop operators remain $1/2$ BPS also for $k=1$ and $2$, preserving sixteen of the thirty-two supercharges ({\em i.e.} four out of the eight extra supercharges). In the Abelian theory, we expect the $1/3$ BPS vortex loop operators, which preserve eight supercharges, to be invariant under {\em all}  the extra eight supercharges that exist for $k=1,2$, and to become $1/2$ BPS. This can be motivated by the fact that with $\cN=8$ supersymmetry the holomorphic and anti-holomorphic fields $C^I$ and $C_I^\dagger$ are in the same multiplet of the $SO(8)$ R-symmetry group. The $1/3$ BPS scale invariant loop operator \eqn{singuthird} is such that the anti-holomorphic field
\be
  C^\dagger_2=\frac{\bar\beta_2}{\sqrt{z}}\propto C^1\,.
\label{1/3to1/2}
\ee
With the extra R-symmetry generators the field $C^\dagger_2$ can be rotated then into $C^1$ and we end up with the same configuration as the $1/2$ BPS operator.

As we point out below, in the non-Abelian theory there will be cases when the $1/3$ BPS vortex loop operators  will have enhanced supersymmetry for $k=1,2$ (when all the vortices are proportional to each-other), and other cases when they do not and they preserve only eight supercharges, which is $1/4$ of the total thirty-two.

\subsubsection*{$\bullet\quad  ${$1/6$ BPS Loop Operators}}

Imposing that  the operator $V_C$ leaves invariant only one of the chiral 
Poincar\'e supersymmetry transformations --- that with label $\epsilon^+_{12}$ 
(and by equation (\ref{complex}) also the anti-chiral one 
$\epsilon^-_{34}$) gives rise to the following  BPS equations
\be
D_{\bar z}C^1=D_{\bar{z}}C^2=D_{z}C^3=D_{{z}}C^4=0\,,\qquad 
D_{t}C^1=D_{t}C^2=D_{t}C^3=D_{t}C^4=0 \,.
\label{BPSb}
\ee

The   static solutions  of these equations are characterized by three holomorphic functions $f_1(z)$, $f_2(z)$, $g(z)$ and two antiholomorphic ones $f_3(\bar{z})$, $f_4(\bar{z})$ all with singularities at $z=0$  
\be
C^1=f_1(z)\qquad C^2=f_2(z)\qquad  C^3=f_3(\bar{z})\qquad C^4=f_4(\bar{z})\qquad A^+_z=g(z)\qquad A^-=0\,.
\ee
Moreover, by the equation of motion for the gauge fields (\ref{eomnonab}), an electric field for $A^+$ must be turned on 
\be
A_t^+=-{4\pi}\left(|f_1|^2+|f_2|^2-|f_3|^2-|f_4|^2\right).
\ee

If we further demand that the singularity produced by $V_C$ is scale invariant, which means it will also preserve the superconformal transformations labeled by $\eta^+_{12}$ and $\eta^-_{34}$, then the form of the singularity is fixed to be
\be
C^1={\beta_1\over \sqrt{z}}\qquad 
C^2={\beta_2\over \sqrt{{z}}}\qquad 
C^3={\beta_3\over \sqrt{\bar{z}}}\qquad 
C^4={\beta_4\over \sqrt{{\bar{z}}}}\qquad 
A^+_z=-i{\alpha\over 2kz}\,.
\label{singusixth}
\ee 
The bosonic symmetries preserved by the 1/6 BPS operators are $SU(1,1)\times U(1)_{\hat d}$,  where $U(1)_{\hat d}$ is a diagonal combination  of a  space-time and $R$-symmetry.\footnote{It is the diagonal sum of $U(1)_l \subset SO(2,3)$ and a $U(1)_{\hat R}\subset SU(4)$ under which $C^I$ have charges $(1,1,-1,-1)$.} 

In the Abelian theory, however, the singularity  given by (\ref{singusixth}) has enhanced symmetry, as $C^1$ and $C^2$ are proportional to each-other, as are $C^3$ and $C^4$. Therefore (\ref{singusixth}) can be transformed into 
\eqn{singuthird} by an $SU(4)$ transformation and is thus $1/3$ BPS. But as we shall see in the analysis for the $U(N)\times U(N)$ theory, in that case it is possible to take $C^1\,/\hskip-3.7mm\propto C^2$ and $C^3\,/\hskip-3.7mm\propto C^4$ and the operators are genuinely ${1/ 6}$ BPS, and are invariant under an $SU(1,1|1)$ subgroup of the $OSp(6|4)$ symmetry of the theory.

\subsubsection{Circular Loop Operators}
\label{sec-circle}

The codimension two singularities we have found as solutions to the BPS equations for the case when the loop operator $V_C$ is supported on a line $C=\bR\subset\bR^3$ can be used to construct   supersymmetric loop operators $V_C$ supported on an arbitrary curve $C\subset\bR^3$. Such a loop operator will be described locally by singularities similar to those we have found for the straight line, but where now the coordinates $(z,\bar{z})$ are interpreted as local coordinates on the  normal bundle of  $C$. In this paper we focus   on supersymmetric loop operators preserving some conformal symmetries.

The only curves in $\bR^3$, other than straight lines, invariant under conformal transformations are circles. Therefore there exist supersymmetric loop operators $V_C$  supported on a circle $C=S^1\subset\bR^3$ which preserve the same superalgebra as the loop operator $V_C$  supported on a line $C=\bR\subset\bR^3$. Since an $S^1$ is related by a global conformal transformation to the line $\bR$, the two curves are $SU(1,1)\times U(1)$ invariant. The operator 
$V_C$ for $C=S^1$ also preserves the same number of supercharges as the corresponding operator for the straight line, 
but in the case of the circle, it is not invariant separately under the Poincar\'e and conformal supercharges, rather under linear combinations of the two.

To construct $V_{S^1}$ explicitly, we consider an $S^1\subset\bR^3$ of radius $a$ located at $t=0, |z|^2=a^2$ in the coordinate system
\be
ds^2=dt^2+dr^2+r^2\,d\psi^2\,,
\ee
then the singularities produced for the scale invariant $1/2,1/3$ and $1/6$ BPS circular loop operators can be obtained from the singularities of the corresponding BPS line operators \eqn{singuconf}, \eqn{singuthird}, \eqn{singusixth} by making the following replacement\footnote{This is most easily derived by a Weyl transformation from $\bR^3$ to $AdS_2\times S^1$, which we discuss below, see \eqn{weyl-trans}.}
\be
z\rightarrow \tilde{r}e^{i\phi}\qquad \bar{z}\rightarrow \tilde{r}e^{-i\phi},
\label{transform}
\ee 
where 
\be
\tilde r^2=\frac{(r^2+t^2-a^2)^2+4a^2t^2}{4a^2}\,
\label{distcircle}
\ee
is the conformal invariant distance from the circle and $\phi$ is the angular coordinate 
defined by
\be
\sin\phi={t\over \tilde{r}}\,.
\ee

\subsubsection{Loop Operators as Vacua   of ${\cal N}=6$ Chern-Simons Theory on $AdS_2\times S^1$}
\label{sec-CS-AdS2}

An alternative way to study loop operators $V_C$ for $C=\bR$ and $C=S^1$ is to study the gauge theory on $AdS_2\times S^1$ instead of $\bR^3$. The analysis in $AdS_2\times S^1$ has the advantage that the symmetries of the scale invariant operators  are realized as isometries of $AdS_2\times S^1$, and not as conformal symmetries. When the gauge theory is studied in $AdS_2\times S^1$, the symmetries of $V_C$ are made manifest.

The only modification to the bosonic Lagrangian of the theory  (\ref{ABJM}) beyond replacing the flat metric by the $AdS_2\times S^1$ metric is the addition of a conformal coupling for the scalars 
\be
{{\cal L}_\text{conf}}=-k\,{R^{(3)}\over 8} \mathrm{Tr}\,  C_I^{\dagger} C^I\,,
\ee
where $R^{(3)}$ is the scalar curvature of the background metric, which for unit-radius $AdS_2\times S^1$ is $R^{(3)}=-2$.

In this formulation, loop operators $V_C$ are given by $SU(1,1)$ invariant  vacua of the theory. The equation that needs to be solved for each scalar is%
\footnote{Note that there is an alternative formulation of these solutions (also in the flat-space description), where the phase of $C^I$ is absorbed by a singular gauge transformation with $A^-_\phi=\pm\frac{1}{2}$. After this transformation the scalar fields are single-valued, but there is a non-integer holonomy around the $\phi$ circle.}
\be
D_{\phi}C^I\mp {i\over 2} C^I=0
\qquad\Longrightarrow\qquad C^I=\beta_I\,e^{\pm{i\over 2}\phi}\,,
\label{vacuum}
\ee
where $\phi$ is the coordinate parametrizing the $S^1$ in $AdS_2\times S^1$ and $\beta_I$ are constants. The choice of sign in the phase is related to the choice of a holomorphic or antiholomorphic field in $\bR^3$. Similarly to the analysis in $\bR^3$, the equation of motion for the gauge field forces that we turn on an electric field proportional to the volume form of $AdS_2$
\be
F^+\propto \Omega_{AdS_2}\,.
\ee
As in the flat-space formulation, the equations of motion allow us to turn on an extra gauge field
\be
A^+_\phi=\frac{\alpha}{k}\,.
\ee

In the $AdS_2\times S^1$ formulation, the operator $V_C$  is supported at the conformal boundary of $AdS_2\times S^1$. For the case of $C=\bR$ we must consider $AdS_2$ in Poincar\'e coordinates while for $C=S^1$ we must consider $AdS_2$ in global coordinates. In this language, loop operators $V_C$ are determined by smooth boundary conditions at asymptotic infinity of $AdS_2\times S^1$ instead of as  singularities in the interior of $\bR^3$.

To see the relation to the vortex loop operators on $\bR^3$ we write 
the metric on $\bR^3$ as a Weyl transformation of the metric on $AdS_2\times S^1$
\be
ds^2_{\bR^3}=\omega^2 ds^2_{AdS_2\times S^1}\,.
\ee
The conformal factor in the transformation between the two metrics, $\omega$, will give the scalars and the gauge field in $\bR^3$ the requisite singularity
\be
C^I|_{\bR^3}={C^I|_{AdS_2\times S^1}\over\sqrt\omega}\qquad   A^{\pm}|_{\bR^3}=A^{\pm}|_{AdS_2\times S^1}\,,
\ee
as $C^I$ has Weyl weight one-half and $A^\pm$ (in form notation) has weight zero.

In the case of the line in $\bR^3$, it is located at $r=0$ in the coordinate system
\be
ds_{\bR^3}=dt^2+dr^2+r^2d\phi^2=r^2\left[{dt^2+dr^2\over r^2}+d\phi^2 \right],
\ee
where $[\cdots]$ is the $AdS_2\times S^1$ metric in Poincar\'e coordinates and $\omega=r$. Combining this Weyl factor and the $AdS_2\times S^1$ vacuum configuration (\ref{vacuum}), we identify $z=re^{i\phi}$ and recover the singularities produced by $V_C$ in $\bR^3$ for $C=\bR$ \eqn{singuconf}.

The circle in $\bR^3$  is located at $r=a$ and $t=0$ in    
\be
ds_{\bR^3}=dt^2+dr^2+r^2d\psi^2={\tilde r}^2\left[d\rho^2+\sinh^2\rho\, d\psi^2+d\phi^2 \right]\,
\ee
where $[\cdots]$ is the $AdS_2\times S^1$ metric in global coordinates and $\omega=\tilde{r}$, where
\be
\begin{aligned}
\tilde r^2&=\frac{(r^2+t^2-a^2)^2+4a^2t^2}{4a^2}
=\frac{a^2}{(\cosh\rho-\cos\phi)^2}
\\
r&=\tilde r\sinh\rho\,\qquad\qquad
t=\tilde r\sin\phi\,.
\end{aligned}
\label{weyl-trans}
\ee
Combining this Weyl factor and the $AdS_2\times S^1$ vacuum configuration (\ref{vacuum}), we get the singularities produced by $V_C$ in $\bR^3$ for $C=S^1$ (\ref{transform}).

The $AdS_2\times S^1$ formulation of $V_C$ makes manifest that the singularities we constructed in $\bR^3$ are $SU(1,1)$ invariant,  since in this formulation the scalar fields have no dependence on the $AdS_2$ coordinates and the required electric field is proportional to the $AdS_2$ volume form.

The $AdS_2\times S^1$ formulation of $V_C$ is also useful in finding the bulk, holographic description of these operators in  $AdS_4\times S^7/\bZ_k$. In Section~\ref{sec-M} we choose to work in a coordinate system where the $AdS_4$ metric is foliated by $AdS_2\times S^1$ slices, and in this foliation the boundary ${\cal N}=6$ Chern-Simons theory is defined on $AdS_2\times S^1$.

\subsection{Loop Operators in the $U(N)\times U(N)$ Theory}
\label{sec-nonabelian}

We now extend the construction of supersymmetric loop operators $V_C$ to the
non-Abelian theory. For simplicity, we will focus on the operators that are scale invariant, that is operators 
defined by a scale invariant codimension two singularity. Moreover, we will write explicitly the singularity for the case when $C=\bR$. One can then construct the singularity when $C=S^1$ by using the transformation (\ref{transform}).
The corresponding description of the loop operators when the theory is on $AdS_2\times S^1$ proceeds in exactly in the same manner as in Section~\ref{sec-CS-AdS2}.

The loop operator $V_C$ in the $U(N)\times U(N)$ theory will have a specified singularity for one or more of the scalar fields along the curve $C$. This singularity will in general break the $U(N)\times U(N)$ gauge symmetry in the vicinity of the loop operator $V_C$ to the subgroup
\be
L=U(N_0)\times U(N_0)\times U(N_1)\times\cdots \times U(N_M)\,, 
\label{levi}
\ee
where $\sum_{l=0}^M N_l=N$. Therefore, the first piece of data that must be specified is a collection of integers $(N_0,\cdots,N_M)$ that form a partition of $N$. Note that for the first number ---  $N_0$ --- there are two factors of $U(N_0)$, while for all the others just one. The reason is that in this first block none of the scalar fields will get a VEV and the gauge symmetry is not broken to the diagonal subgroup.
 
The precise definition of the loop operator $V_C$ is as follows. First we specify the unbroken gauge symmetry as in \eqn{levi} and an $L$-invariant singularity produced by  $V_C$, on which we elaborate below. Then the operator $V_C$ is defined by the path integral over all smooth field configurations with the same $L$-invariant  singularity near $C$. In performing the path integral, one must mod out by the gauge transformations that take values in $L\subset U(N)\times U(N)$ when restricted to $C$.

We now consider the various BPS loop operators in the $U(N)\times U(N)$ theory.

\subsubsection*{$\bullet\quad ${$1/2$ BPS Loop Operators}}

In the non-Abelian theory, the BPS equations describing a 1/2 BPS loop operator $V_C$ preserving the supercharges parameterized by \eqn{six} are still given by 
\be
D_{\bar z}C^1=0\qquad D_t C^1=0\,,
\ee
where  now
\be
DC^1=dC^1-i(AC^1-C^1\hat{A})\,,
\ee
and $C^2$, $C^3$ and $C^4$ are constants. These equations  must be supplemented with the equations of motion for the gauge fields (\ref{eomnonab}).

Any static solution of this equation can be diagonalized by a $U(N)\times U(N)$ transformation. Focusing on the conformally invariant solutions, $C^2=C^3=C^4=0$ and 
 the singularity of the complex scalar field $C^1$  is then given by
\be
C^1=\frac{1}{\sqrt{z}}
\begin{pmatrix}
0\otimes 1_{N_0}
&0 &\cdots & 0\cr
0& \beta^{(1)}\otimes1_{N_1}&\cdots & 0 \cr
\vdots &\vdots &\ddots&\vdots\cr
0&0&\cdots&\beta^{(M)}\otimes1_{N_M}
\end{pmatrix}.
\label{holomorphic}
\ee
The scalar field acquires a $U(N_0)^2\times U(N_1)\times\cdots\times U(N_M)$ invariant singularity, labeled by $M$ real positive parameters $(\beta^{(1)},\cdots,\beta^{(M)})$, where we have removed the phases of all the $\beta^{(l)}$ by perfoming a $U(1)^M$ gauge transformation.

As in the $U(1)\times U(1)$ theory, we consider solutions to the BPS equations where
\be
A=\hat{A}\,.
\ee 
We can therefore identify the gauge indices of the two gauge groups and define again $A^+=A+\hat A$ (and $A^-=0$). 
The first BPS equation, together with (\ref{holomorphic}) implies that 
\be
[C,A^+_z]=0\,.
\ee
Therefore, $A^+_z$ is given by an arbitrary  diagonal matrix. For a $U(N_0)^2\times U(N_1)\times\cdots U(N_M)$ invariant singularity, the diagonal gauge field  produced by $V_C$ takes the following form
\be
A^+_z=-\frac{i}{2kz}
\begin{pmatrix}
0\otimes 1_{N_0}
&0 &\cdots & 0\cr
0& \alpha^{(1)}\otimes1_{N_1}&\cdots & 0 \cr
\vdots &\vdots &\ddots&\vdots\cr
0&0&\cdots&\alpha^{(M)}\otimes1_{N_M}
\end{pmatrix}\,.
\label{holomorphicgauge}
\ee
The parameters $\alpha_l$ are defined with unit period.  The equation of motion for the gauge fields requires that we turn on an electric field for the $A^+$ gauge field, which in complete analogy with the Abelian case can be represented by the vector potential
\be
A^+_t=-4\pi C^1\,C_1^\dagger\,.
\ee

In summary, a $1/2$ BPS loop operator $V_C$ with $L=U(N_0)^2\times U(N_1)\times\cdots U(N_M)$ is labeled by $2M$ parameters $(\alpha^{(l)},\beta^{(l)})$, where $l=1,\cdots,M$.

\subsubsection*{$\bullet\quad ${$1/3$ BPS Loop Operators}}

In the non-Abelian theory, the BPS equations describing the ${1/3}$ BPS loop operators $V_C$ preserving the supercharges parametrized by \eqn{four} are  given by
\be
D_{\bar z}C^1= D_{z}C^2= 0,\quad
D_t C^1=D_t C^2=0,\quad
C^1C_1^\dagger C_2= C_2 C_1^\dagger C^1,\quad
C^2C_2^\dagger C_1= C_1 C_2^\dagger C^2,
\label{BPSthirdNA}
\ee
with constant $C^3$ and $C^4$. In addition we have to impose the equations of motion for the gauge fields (\ref{eomnonab}). 

As in the ${1/2}$ BPS case, taking the conformally invariant case, $C^3=C^4=0$ and we can diagonalize $C^1$ by a $U(N)\times U(N)$ transformation 
\be
C^1=\frac{1}{\sqrt{z}}
\begin{pmatrix}
0\otimes 1_{N_0}
&0 &\cdots & 0\cr
0& \beta_1^{(1)}\otimes1_{N_1}&\cdots & 0 \cr
\vdots &\vdots &\ddots&\vdots\cr
0&0&\cdots&\beta_1^{(M)}\otimes1_{N_M}
\end{pmatrix}\,.
\label{holomorphic3}
\ee
The last two equations in (\ref{BPSthirdNA}) further imply that the matrix $C^2$ can be simultaneously diagonalized 
so the second scalar field develops the following singularity
\be
C^2=\frac{1}{\sqrt{\bar{z}}}
\begin{pmatrix}
0\otimes 1_{N_0}
&0 &\cdots & 0\cr
0& \beta_2^{(1)}\otimes1_{N_1}&\cdots & 0 \cr
\vdots &\vdots &\ddots&\vdots\cr
0&0&\cdots&\beta_2^{(M)}\otimes1_{N_M}
\end{pmatrix}\,.
\label{holomorphic4}
\ee
The singularities  arising from the scalars are labeled by $2M$ complex parameters $(\beta_1^{(l)},\beta_2^{(l)})$ subject to the relation $(\beta_1^{(l)},\beta_2^{(l)})\simeq e^{i\theta_l}(\beta_1^{(l)},\beta_2^{(l)})$  for $l=1,\cdots M$,
thus resulting in $3M$ real parameters.

The singularity for the gauge field is unmodified from the $1/2$ BPS  case and is given by (\ref{holomorphicgauge}). As in the Abelian case an electric field for $A^+$ must also be turned on and is completely determined by $C^1$ and $C^2$ 
\be
A_t=-4\pi\left(C^1\,C_1^\dagger-C^2\,C_2^\dagger\right).
\ee

We mentioned for the theory with $U(1)\times U(1)$ gauge symmetry that in the case of $k=1,2$, where the theory is expected to have enhanced $\cN=8$ supersymmetry, the supersymmetry of the $1/3$ BPS vortex is enlarged by eight more supercharges to a total of sixteen, so it becomes $1/2$ BPS. Does the same happen for the non-Abelian vortex?

The argument from the $U(1)\times U(1)$ theory can be carried over to our discussion here, only that while there equation \eqn{1/3to1/2} was automatically satisfied, now it will have to be imposed as an extra constraint. Therefore the $1/3$ BPS vortex will have enhanced supersymmetry for $k=1,2$ if and only if the parameters $\beta_1^{(l)}$ and $\beta_2^{(l)}$ are such that the matrices $C^1$ and $C^\dagger_2$ are proportional to each-other.

In summary, a ${1/3}$ BPS loop operator $V_C$ with $L=U(N_0)^2\times U(N_1)\times\cdots U(N_M)$ is labeled by $4M$ real parameters $(\alpha^{(l)},\beta_1^{(l)},\beta_2^{(l)})/U(1)^M$, where $l=1,\cdots,M$. The ones that are $1/2$ BPS for $k=1,2$ are labeled by $2M+2$ real parameters, $(\alpha^{(l)},|\beta_1^{(l)}|)$ and the constant ratio between $C^1$ and $C_2^\dagger$.

\subsubsection*{$\bullet\quad ${$1/6$ BPS Loop Operators}}

In the non-Abelian theory, the BPS equations describing the ${1/6}$ BPS loop operators 
$V_C$ invariant under the supersymmetry transformations with parameters 
$\epsilon^+_{12}$ and $\epsilon^-_{34}$ are  given by 
\be
D_{\bar z}C^1= D_{\bar{z}}C^2= D_{z}C^3= D_{z}C^4=0\,,\qquad  D_t C^1=D_t C^2=D_t C^3=D_t C^4=0\,.
\label{BPSthirdNA1}
\ee
The scalars fields $C^I$ must also satisfy certain matrix constraints analogous to those in (\ref{BPSthirdNA}), which are solved when all four matrices are diagonal. 
These equations  must be supplemented with the equations of motion for the gauge fields (\ref{eomnonab}). 

The solutions to \eqn{BPSthirdNA1} preserving conformal invariance are of the form \eqn{holomorphic3} for the scalars $C^1$ and $C^2$ and \eqn{holomorphic4} for $C^3$ and $C^4$. Taking the indices $(\tilde I, \hat I)$ to label $C^1,C^2$ and $C^3,C^4$ respectively, the singularities induced on the scalar fields by the ${1/6}$ BPS loop operators $V_C$ are given by
\be
C^{\tilde I}=\frac{1}{\sqrt{z}}
\begin{pmatrix}
0\otimes 1_{N_0}
&0 &\cdots & 0\cr
0& \beta_{\tilde I}^{(1)}\otimes1_{N_1}&\cdots & 0 \cr
\vdots &\vdots &\ddots&\vdots\cr
0&0&\cdots&\beta_{\tilde I}^{(M)}\otimes1_{N_M}
\end{pmatrix}\,
\label{holomorphic6}
\ee
and 
\be
C^{\hat I}=\frac{1}{\sqrt{\bar{z}}}
\begin{pmatrix}
0\otimes 1_{N_0}
&0 &\cdots & 0\cr
0& \beta_{\hat I}^{(1)}\otimes1_{N_1}&\cdots & 0 \cr
\vdots &\vdots &\ddots&\vdots\cr
0&0&\cdots&\beta_{\hat I}^{(M)}\otimes1_{N_M}
\end{pmatrix}\,.
\label{holomorphic6b}
\ee
The singularities  arising from the scalars are labeled by $4M$ complex parameters $\beta_I^{(l)}$ subject to the relation $(\beta_{I}^{(l)})\simeq e^{i\theta_l}(\beta_{I}^{(l)})$  for $l=1,\cdots M$ and $I=1,\ldots,4$, thus resulting in $7M$ real parameters.

As in the ${1/3}$ BPS case,  the singularity for the gauge field is unmodified from the ${1/2}$ BPS  case and is given by (\ref{holomorphicgauge}). An electric field for $A^+$ must also be excited
\be
A_t=-4\pi\left(C^1\,C_1^\dagger+C^2\,C_2^\dagger
-C^3\,C_3^\dagger-C^4\,C_4^\dagger\right).
\ee

In summary, a ${1/6}$ BPS loop operator $V_C$ with $L=U(N_0)^2\times U(N_1)\times\cdots U(N_M)$ is labeled by $8M$ parameters $(\alpha^{(l)},\beta_a^{(l)},\beta_{a'}^{(l)})/U(1)^M$, where $l=1,\cdots,M$. Some degenerate cases will preserve more than four supercharges (for example when $M=1$), or have enhanced supersymmetry when $k=1,2$.

\subsubsection*{$\bullet\quad ${Vortices at odd level $k$}}

As mentioned above, in the case of the theory with $U(1)\times U(1)$ gauge 
symmetry, the vortices are a good gauge theory background only for the theory 
with even level $k$. For odd level there are gauge-invariant local observables 
which are not single-valued when encircling these vortices. The same is true 
for the construction we presented here in the non-Abelian theory. 
We would like to comment here about a modification of this construction which 
applies also for odd $k$, inspired by a similar construction for surface operators 
in $\cN=4$ SYM in four dimensions of Koh and Yamaguchi \cite{Koh-Yamaguchi}.

For this modification one needs to take all the integers $N_l$ with $1<l\leq M$ to 
be even and then break every $N_l\times N_l$ block in two. The singularity of 
the scalar field \eqn{holomorphic} is then modified such that half of the eigenvalues 
in each block have the opposite sign
\be
C^1=\frac{1}{\sqrt{z}}
\begin{pmatrix}
0\otimes 1_{N_0}
&0 &\cdots & 0\cr
0&\begin{pmatrix}\beta^{(1)}&0\cr0&-\beta^{(1)}\end{pmatrix}\otimes1_{N_1/2}&\cdots & 0 \cr
\vdots &\vdots &\ddots&\vdots\cr
0&0&\cdots&\begin{pmatrix}\beta^{(M)}&0\cr0&-\beta^{(M)}\end{pmatrix}\otimes1_{N_M/2}
\end{pmatrix},
\ee
which can also be written as
\be
C^1=\frac{1}{\sqrt{z}}
\begin{pmatrix}
0\otimes 1_{N_0}
&0 &\cdots & 0\cr
0&\beta^{(1)}\,\sigma_3\otimes1_{N_1/2}&\cdots & 0 \cr
\vdots &\vdots &\ddots&\vdots\cr
0&0&\cdots&\beta^{(M)}\,\sigma_3\otimes1_{N_M/2}
\end{pmatrix},
\label{holomorphic/2}
\ee
with $\sigma_3$ a Pauli matrix.

So far it seems like a vortex with gauge symmetry broken to 
$L=U(N_0)^2\times U(N_1/2)^2\times\cdots U(N_M/2)^2$, but the
novel feature proposed in \cite{Koh-Yamaguchi} is to add a non-trivial 
gauge twist around the vortex, which breaks the symmetry to 
$L=U(N_0)^2\times U(N_1/2)\times\cdots U(N_M/2)$. Instead of 
\eqn{holomorphicgauge} we take the holonomy to be
\be
\exp i\oint A^+_z\,dz=
\begin{pmatrix}
1_{N_0}
&0 &\cdots & 0\cr
0& e^{i\pi\alpha^{(1)}/k}\,\sigma_1\otimes1_{N_1/2}&\cdots & 0 \cr
\vdots &\vdots &\ddots&\vdots\cr
0&0&\cdots&e^{i\pi\alpha^{(M)}/k}\,\sigma_1\otimes1_{N_M/2}
\end{pmatrix}\,.
\label{twist}
\ee
Then, when going around the vortex, the Pauli matrices $\sigma_1$ permute the 
pairs of eigenvalues in \eqn{holomorphic/2}, so as opposed to the general 
case \eqn{holomorphic}, this construction is in fact single-valued around the 
vortex. Such configurations are perfectly good backgrounds 
for the gauge theory also for odd $k$, even in the presence of operators of 
the form $C^k$.

To summarize, for odd $k$ the general $1/2$ BPS vortex loop operators has 
unbroken gauge symmetry $L=U(N_0)^2\times U(N_1/2)\times\cdots U(N_M/2)$ 
and is labeled by $2M$ parameters $(\alpha^{(l)},\beta^{(l)})$, where $l=1,\cdots,M$. 
Similar constructions apply also for $1/3$ BPS and $1/6$ BPS vortex loops.

\subsection{Vacuum Expectation Value}
\label{sec-vev-CS}

Conformal invariance implies that the one point function of a local operator  must vanish. This need not be the case for non-local operators, and the expectation value of non-local operators have played an important role as order parameters of phases of gauge theories. 

Our first task will be to compute, in the semiclassical approximation, the expectation value of the BPS disorder loop operators that we have constructed. This is achieved by evaluating the classical Euclidean action of  ${\cal N}=6$ Chern-Simons theory on the field configuration produced by the operator $V_C$
\be
\langle V_C \rangle =\exp\left(-S_\text{class.}\right)\,.
\ee 

This computation is easily performed by considering the description of a loop operator as a vacuum state of the theory on $AdS_2\times S^1$. The relevant part of the Euclidean Lagrangian is
\be
{\cal L}= k\,\hbox{Tr}\left(D_\mu C^\dagger_ID^\mu C^I+{R^{(3)}\over 8}C_I^\dagger C^I\right)\,,
\ee
where as mentioned earlier $R^{(3)}=-2$ for $AdS_2\times S^1$. We have not included the Chern-Simons terms for the gauge fields as they  trivially vanish when evaluated on the gauge field configuration excited by  $V_C$. 
Since $C^I=C_0^Ie^{\pm {i\over 2}\phi}$, where $C_0^I$ is a constant diagonal matrix made of the parameters $\beta_I^{(l)}$,  we have that $|DC^I|^2=|dC^I|^2={1\over 4}|C^I|^2$, which cancels the conformal coupling of the scalars.  Therefore the on-shell action vanishes and 
\be
\langle V_C \rangle =1
\ee
in the semiclassical approximation. We note that $\langle V_C \rangle =1$ both for $C=\bR$ and $C=S^1$ as the vanishing of the on-shell action holds for both Poincar\'e and global  $AdS_2$.

The same conclusion can be reached by evaluating the on-shell action for the singularity produced by $V_C$ in $\bR^3$. Care must be taken, however, to ensure that the action has a well defined variational principle and that the
boundary action vanishes when evaluated on the 
singularity.\footnote{See \cite{Drukker:2008wr} for the corresponding analysis of the on-shell action for surface operators in ${\cal N}=4$ SYM.} This requires adding a boundary term to the action in (\ref{ABJM}), whose net effect is to cancel the bulk term when evaluated on-shell.

\subsection{Correlator with Local Operators}
\label{sec-local-CS}

In this section we calculate various correlators involving the BPS loop operators we found in the previous section. 
We calculate the correlator of a BPS loop operator with chiral primary operators and the stress tensor in ${\cal N}=6$ Chern-Simons theory.
See \cite{Drukker:2008wr} for a closely related discussion in the context of disorder surface operators in four dimensional ${\cal N}=4$ SYM.

In the semiclassical approximation, the correlation function of a loop operator $V_C$ and a local operator ${\cal O}$ in ${\cal N}=6$ Chern-Simons theory in $\bR^3$ is found by evaluating the operator ${\cal O}$ in the background field that the loop operator produces
\be
{\langle V_C\cdot {\cal O}\rangle\over \langle V_C \rangle}= {\cal O}|_\text{loop}\,.
\ee

Conformal Ward identities constrain the form of the correlator of $V_C$ with a   local operator ${\cal O}$. When $C=\bR$ the dependence of the correlator with a dimension $\Delta$ scalar operator on the distance $r$ is given by
\be
{\langle V_C\cdot {\cal O}\rangle\over \langle V_C \rangle}={c_{\cal O}\over r^\Delta}\,.
\ee 
The correlator is captured by the coefficient $c_{\cal O}$, which depends on the charges of the operator, the 't Hooft coupling $\lambda$ and $N$. 
 When $C=S^1$ the correlator is given by 
\be
{\langle V_C\cdot {\cal O}\rangle\over \langle V_C \rangle}={c_{\cal O}\over {\tilde r}^\Delta}\,,
\ee 
where $\tilde{r}$ is defined in (\ref{distcircle}).
In the calculation below we determine the value of $c_{\cal O}$ in the semiclassical approximation.

We now proceed to compute --- in the semiclassical approximation --- the correlator between a 1/2 BPS loop operator $V_C$ and 
the simplest chiral primary operators   in ${\cal N}=6$ Chern-Simons theory. The operators we consider here, ${\cal O}^A_\Delta$ of conformal dimension $\Delta$, transform in the $[\Delta,0,\Delta]$ representation of the $SU(4)$ R-symmetry group \cite{Aharony:2008ug}.%
\footnote{These operators carry zero $U(1)_B$ ``baryonic'' charge and have an equal number of $C$ and $C^\dagger$ fields. We comment below on the more general operators.} 
The expression for the unit normalized chiral primary operators in the planar approximation is given by
\be
{\cal O}^A_\Delta={(4\pi)^\Delta\over \lambda^{\Delta}\sqrt{\Delta}}C^{(A)}{}^{J_1\cdots J_\Delta}_{I_1\cdots I_\Delta}\, \hbox{Tr}\left(  C^{I_1}C^\dagger_{J_1}\cdots C^{I_\Delta}C^\dagger_{J_\Delta}\right)\,,
\label{general-local}
\ee
where $C^{(A)}{}^{J_1\cdots J_\Delta}_{I_1\cdots I_\Delta}$ is a totally symmetric tensor in $I_1\cdots I_\Delta$  and 
$J_1\cdots J_\Delta$ which vanishes when the trace is taken between any $I$ and $J$ index. 
The tensor $C^{(A)}{}^{J_1\cdots J_\Delta}_{I_1\cdots I_\Delta}$ is normalized by
\be
C^{(A)}{}^{J_1\cdots J_\Delta}_{I_1\cdots I_\Delta}\bar{C}^{(B)}{}^{I_1\cdots I_\Delta}_{J_1\cdots J_\Delta}=\delta^{AB}\,.
\ee
This guarantees that the operator ${\cal O}$, is unit normalized as%
\footnote{The propagator for the scalar fields is given by $ 
\vev{C^I{}^i_{\hat i}(x) C^\dagger_J{}^{\hat j}_j(y)}=\frac{1}{4\pi k}{1\over |x-y|}\delta^I_J\delta^i_j\delta^{\hat j}_{\hat i}$.}
\be
\langle {\cal O}(x) \bar{{\cal O}}(y)\rangle={1\over |x-y|^{2\Delta}}\,.
\label{unidad}
\ee

Since the $1/2$ BPS loop operators $V_C$ are $SU(3)$ invariant, the chiral primary operators that have a non-vanishing correlator with $V_C$ are the $SU(3)$ invariant ones. In the decomposition of the $[\Delta,0,\Delta]$ representation of $SU(4)$ under the maximal $SU(3)\times U(1)_R$ subgroup, there is a unique operator for each $\Delta$ which is an   $SU(3)$ singlet and which has a non-trivial correlator with  $V_C$. We label this operator $\cO_{\Delta,0}$.\footnote{The subscript $0$ is used to indicate that these operators have vanishing ``baryonic" charge.} 
For a detailed discussion see Appendix~\ref{app-harmonics}.

The $SU(3)$ invariant chiral primary operators  in (\ref{general-local})  are related to the spherical harmonics on $S^7$ by \eqn{harmon}, \eqn{normalized}
\be
C^\Delta{}^{J_1\cdots J_\Delta}_{I_1\cdots I_\Delta} 
w^{I_1}\cdots w^{I_\Delta}\bar w_{J_1}\cdots\bar w_{J_\Delta}=\frac{\sqrt{2}\,\Delta!}{\sqrt{(2\Delta+2)!}}\,P_\Delta^{(0,2)}(\cos\vartheta_1)\,,
\label{norm2}
\ee
where $P_n^{(\alpha,\beta)}$ is a Jacobi polynomial and $w^I$ are  coordinates in $\bC^4$ defined in \eqn{ws}, which get identified with the fields $C^I$. The argument of the polynomial is given by $\cos\vartheta_1=1-2|w^1|^2$.

The explicit form of the low dimension operators are given by \eqn{Os}
\be
\begin{aligned}
\cO_{1,0}&=\frac{2\pi}{\sqrt{3}\lambda}\Tr\Big[C^IC^\dagger_I-4C^1C_1^\dagger\Big],\\
\cO_{2,0}&=\frac{8\pi^2}{3\sqrt{5}\lambda^2}
\Tr\Big[(C^IC_I^\dagger)^2-10C^IC_I^\dagger\,C^1C_1^\dagger+15(C^1C_1^\dagger)^2\Big],\\
\cO_{3,0}&=\frac{16\pi^3}{\sqrt{105}\lambda^3}
\Tr\Big[(C^IC_I^\dagger)^3-18(C^IC_I^\dagger)^2\,(C^1C_1^\dagger)+
63(C^IC_I^\dagger)\,(C^1C_1^\dagger)^2-56(C^1C_1^\dagger)^3\Big].
\end{aligned}
\label{opss}
\ee
Note that all products of fields should be symmetrized and the index $I$ is summed from 1 to 4.

Evaluating semiclassically the expectation value  of these local operators in the 1/2 BPS vortex loop operator background  amounts to inserting (\ref{holomorphic})   in the expression for the chiral primary operator. Since on-shell $C^2=C^3=C^4=0$, the operator is proportional to $(C^1C^\dagger_1)^\Delta$. Then we plug into the spherical harmonic $\vartheta_1=\pi$ ({\em i.e.} $w^1=1$) which gives
 \be
P_\Delta^{(0,2)}(-1)
=(-1)^\Delta\frac{(\Delta+1)(\Delta+2)}{2}\,.
\ee
From this we find that the correlator between a unit normalized chiral primary operator and a 1/2 BPS vortex loop operator is given by
\be
\frac{\vev{V_C\cdot {\cal O}_{\Delta,0}}}{\vev{ V_C}}={(-1)^\Delta\over |z|^\Delta}
\left(\frac{4\pi}{\lambda}\right)^\Delta
{(\Delta+2)!\over \sqrt{2\Delta (2\Delta+2)!}}\sum_{l=1}^M N_l\, |\beta^{(l)}|^{2\Delta}\,.
\label{pert-correlator}
\ee

This far we have focused on the chiral primary operators with equal number of $C$ and $C^\dagger$ fields. There are other chiral primary operators in the theory that are $SU(3)$ invariant and which carry $U(1)_B$ ``baryonic'' charge, measuring the difference in the number of $C$ and $C^\dagger$ fields. Gauge invariance  in ${\cal N}=6$ Chern-Simons theory at level $k$  restricts the charge of these operators to be $pk$, where $p$ is an integer. The chiral primary operators of this type transform in the $[\Delta\pm \frac{pk}{2},0,\Delta\mp\frac{pk}{2}]$ representation of $SU(4)$ (where $\Delta\geq |pk/2|$). The simplest ones --- those with $\Delta=|pk/2|$ --- can be schematically written (taking $p>0$) as
\be
\cO_{\frac{pk}{2},p}\sim\frac{1}{\lambda^{pk/2}}
(C^I)^{pk}\,,\qquad 
\cO_{\frac{pk}{2},-p}\sim\frac{1}{\lambda^{pk/2}}
(C_I^\dagger)^{pk}\,.
\label{toft}
\ee
Gauge invariance requires that $\pm p$ units of flux are threaded through the $S^2$ surrounding the point where the operator is inserted \cite{Aharony:2008ug}. As before, the correlator of such a  chiral primary operator with 
a vortex loop operator $V_C$  can be computed by inserting the field produced by $V_C$ in (\ref{toft}). This yields%
\footnote{Though it is natural to guess that they will scale like $(\beta^{(l)})^{pk}$, the incomplete understanding of these operators prevents us from determining the proper normalization as well as the detailed dependence on $\beta^{(l)}$.}
\be
\frac{\vev{V_C\cdot \cO_{\frac{pk}{2},p}}}{\vev{ V_C}}
\sim\frac{1}{\lambda^{pk/2}} \frac{1}{z^{pk/2}}\,,
\qquad 
\frac{\vev{V_C\cdot \cO_{\frac{pk}{2},-p}}}{\vev{ V_C}}
\sim \frac{1}{\lambda^{pk/2}}\frac{1}{\bar{z}^{|p|k/2}}\,.
\label{toftcorr}
\ee

As mentioned in Section \ref{secabelia}, some of the  scalar fields $C^I$ are not single-valued when taken around a loop operator $V_C$. Such   discontinuities  in the fields of the Lagrangian are not problematic as long as all the gauge invariant operators of the theory are single valued when encircling  $V_C$. The chiral primary operators 
with $p=0$ (\ref{general-local}) are indeed single valued around $V_C$. On the other hand, it follows from (\ref{toftcorr}) that chiral primary operators with non-vanishing ``baryonic" charge (\ref{toft}) pick up the  phase 
\be
(-1)^{pk}
\label{phase-pk}
\ee
upon encircling $V_C$.  For even level $k$, the operators are single valued, and therefore loop operators  are physical. For odd $k$, however, this simplistic analysis suggests that operators with odd $p$ change sign. This implies that the generic vortex loop operators are unphysical for odd $k$, as they do not give rise to a consistent operator algebra.  An exception is the construction at the end of Section~\ref{sec-nonabelian}, where all the integers $N_l$ with $l=1,\cdots M$ parametrizing the unbroken gauge group \eqn{levi} are even. Then the construction in \eqn{holomorphic/2} and \eqn{twist} interchanges the the eigenvalues $\pm\beta^{(l)}$ upon encircling the vortex, which compensates for the phase \eqn{phase-pk}. In Section \ref{sec-M2}, we will find a bulk counterpart of this statement, where the candidate M2-brane describing  a vortex loop operator exists for odd $k$ only when all the integers $N_l$ are even.

\subsubsection*{$\bullet\quad $  Scaling Weight}

The stress tensor in a CFT plays an important role as it generates   conformal transformations. For non-local operators, one may define the analog of the familiar conformal weight of a local operator from the correlator of the non-local operator with the stress tensor (see e.g \cite{Kapustin:2005py,Drukker:2008wr,Gomis:2008qa}). The form of the correlator of $V_C$ with the stress tensor $T_{\mu\nu}$  when $C=\bR$ is given by
\be
{\langle{T_{00}\cdot V_C}\rangle\over\langle V_C \rangle}
= {h\over r^3},
\qquad
{\langle{T_{ij}\cdot V_C}\rangle\over\langle V_C \rangle}
={h\over r^3}\left[{3n_in_j-2\delta_{ij}}\right],
\qquad
{\langle{T_{0i}\cdot V_C}\rangle}=0\,.
\ee
Here $x^{\mu}=(x^0,x^i)$, where $x^0$ is the coordinate  along
 $C=\bR$  and $n^i=x^i/r$ is the unit normal vector to the straight line.  
The correlator is completely determined up to the function $h$ --- the scaling weight --- which generalizes the notion of conformal dimension of local operators to non-local operators. 

The bosonic contribution to the stress tensor of ${\cal N}=6$ Chern-Simons theory is given by
\begin{eqnarray}
T_{\mu\nu}={2\over \sqrt{g}}{\delta {\cal L}\over \delta g^{\mu\nu}}
=k\,\hbox{Tr}\Big(D_\mu C_I^\dagger D_\nu C^I+D_\mu C^I D_\nu C_I^\dagger -g_{\mu\nu}D_\lambda C_I^\dagger D^\lambda C^I\cr
+{1\over 4}R^{(3)}_{\mu\nu} C_I^\dagger  C^I+{1\over 4}(g_{\mu\nu}D^2-D_\mu D_\nu)C_I^\dagger  C^I-{R^{(3)}\over 8}g_{\mu\nu}C_I^\dagger  C^I -g_{\mu\nu}V_{pot}
\Big)\,,
\end{eqnarray}
where $R^{(3)}$, $R^{(3)}_{\mu\nu}$ denote the scalar curvature and the Ricci tensor of the background on which the gauge theory is defined.

The semiclassical scaling weight for a 1/2, 1/3 and 1/6 BPS  loop operator $V_C$ can be computed semiclassically by evaluating the stress tensor in the background produced by the corresponding loop operator, which yields 
\be
h=-{k\over 4}\sum_{I=1}^{4}\sum_{l=1}^M N_l\, |\beta_I^{(l)}|^2\,.
\ee
This expression is written for the most general $1/6$ BPS vortex loop operator. In the other cases with more supersymmetries, some of the $\beta_I^{(l)}$'s have to be set to zero.

Since the stress tensor is in the same supermultiplet as the $\Delta=1$ chiral primary operator, the correlator
of a vortex loop operator with $T_{\mu\nu}$ and with ${\cal O}_{1,0}$ are related by superconformal Ward identities \cite{Gomis:2008qa}. It would be interesting to study the supercurrent multiplet for ${\cal N}=6$ Chern-Simons theory.

\section{Holographic M-Theory Description}
\label{sec-M}

\subsection{M-Theory on $AdS_4\times S^7/\bZ_k$}

The $\cN=6$ Chern-Simons    theory  with $U(N)_k\times U(N)_{-k}$ gauge group we have been studying is conjectured 
\cite{Aharony:2008ug} to describe  
the low energy limit of the dynamics of $N$ M2-branes on a $\bZ_k$
orbifold of $\bR^8$. Therefore, this theory is expected to provide the holographic description of M-theory with 
 $AdS_4\times S^7/\bZ_k$ boundary conditions. The M-theory background is given
by the following metric and four-form
\be
\begin{aligned}
ds^2&=\frac{R^2}{4}\,ds_{AdS_4}^2+R^2ds_{S^7/\bZ_k}^2\,,\\
F_4&=\frac{3}{8}R^3\,\Omega_{AdS_4}\,,
\end{aligned}
\label{M-metric}
\ee
where $\Omega_{AdS_4}$ is the volume-form on $AdS_4$. 

In order to identify the bulk description of the vortex loop operators $V_C$ found in the previous section, it is
convenient to foliate the bulk $AdS_4$ metric by  $AdS_2\times S^1$  slices, as this makes manifest the symmetries of the dual loop operators.  In this foliation of  $AdS_4$,  the metric in the conformal boundary is that of $AdS_2\times S^1$, where vortex loop operators have a particularly simple description.

In this foliation the $AdS_4$ metric is given by
\be
ds_{AdS_4}^2=du^2+\cosh^2u\, ds^2_{AdS_2}+\sinh^2u\,d\phi^2\,,
\label{metricmaca}
\ee
where $ds^2_{AdS_2}$ is the metric of  $AdS_2$. We can then choose the metric of $AdS_2$ 
in either Poincar\'e or global coordinates 
\begin{align}
ds_{AdS_2}^2&=  {dt^2+dz^2\over z^2}\,,
\label{poincare-AdS2-metric}
\\
ds_{AdS_2}^2&= d\rho^2+\sinh^2\rho\,d\psi^2\,.
\label{global-euclid-metric}
\end{align}
The Poincar\'e coordinates are suitable for describing loop operators supported on $C=\bR$ while global coordinates are suitable when the loop operators are supported on $C=S^1$, 
mirroring the discussion in Section~\ref{sec-CS-AdS2}.
The brane constructions we write down below apply to both choices of $AdS_2$  coordinates. 

To write down the M2-brane action in this background we need also the gauge 
potential for the four-form $F_4$ \eqn{M-metric}. We take
\be
C_3=\frac{1}{8}R^3(\cosh^3u-1)\, \Omega_{AdS_2} \wedge d\phi\,,
\ee
where $\Omega_{AdS_2}$ is the volume form of $AdS_2$.
In principle $C_3$ is defined only up to a gauge choice, but since we
will couple it to branes that 
approach  the boundary of spacetime, one should impose a proper
asymptotic behavior on it. The analog of choosing 
Fefferman-Graham coordinates \cite{Fefferman} near the boundary is to take the three-form to
not have any component in the $du$ direction. Such a prescription indeed 
gave the correct result in $\cN=4$ SYM in four dimensions 
\cite{Drukker:2005kx}.\footnote{See a more detailed discussion in \cite{Drukker:2008wr}.}

We choose a set of coordinates for $S^7/\bZ_k$ defined 
by the embedding of the unit 7-sphere in $\bC^4$ given by
\be
\begin{aligned}
w^1&=\sin\frac{\vartheta_1}{2}\,e^{i\xi_1}\,,\qquad
&w^3&=\cos\frac{\vartheta_1}{2}\cos\frac{\vartheta_2}{2}
\sin\frac{\vartheta_3}{2}\,e^{i\xi_3}\,,\\
w^2&=\cos\frac{\vartheta_1}{2}\sin\frac{\vartheta_2}{2}\,e^{i\xi_2}\,,\qquad
&w^4&=\cos\frac{\vartheta_1}{2}\cos\frac{\vartheta_2}{2}
\cos\frac{\vartheta_3}{2}\,e^{i\xi_4}\,.
\end{aligned}
\label{ws}
\ee
The angles $\vartheta_1$, $\vartheta_2$ and $\vartheta_3$ all range from 0 to $\pi$. The angles $\xi_1$,  $\xi_2$, $\xi_3$ and $\xi_4$ have period $2\pi$ but are subject to the $\bZ_k$ orbifold action
\be
\xi_I\rightarrow \xi_I+2\pi/k\,,
\ee
  identifying $w^I\to e^{2\pi i/k}w^I$.   In this coordinate system the metric on $S^7/\bZ_k$ is given by 
\be
\begin{aligned}
ds^2_{S^7/\bZ^k}=\frac{1}{4}\Bigg[&
d\vartheta_1^2
+4\sin^2\frac{\vartheta_1}{2}d\xi_1^2
+\cos^2\frac{\vartheta_1}{2}\bigg(
d\vartheta_2^2
+4\sin^2\frac{\vartheta_2}{2}d\xi_2^2
\\&\hskip1cm
+\cos^2\frac{\vartheta_2}{2}\bigg(
d\vartheta_3^2
+4\sin^2\frac{\vartheta_3}{2}d\xi_3^2
+4\cos^2\frac{\vartheta_3}{2}d\xi_4^2
\bigg)\bigg)\Bigg]\,.
\end{aligned}
\label{S7-metric}
\ee

The relation between the parameters of the M-theory background and of
the Chern-Simons field theory are
\be
\frac{R^3}{4k}=\pi\sqrt{\frac{2N}{k}}=\pi\sqrt{2\lambda}\,.
\label{radius}
\ee
M-theory should provide a good description of  Chern-Simons theory in the (strong) 't Hooft coupling limit and in the regime $\lambda^{5/2}\gg N^2$. For larger $k$, when $N^2\gg\lambda^{5/2}$, the  perturbative  bulk description is given by Type IIA supergravity on $AdS_4\times CP^3$ \cite{Aharony:2008ug}. Next we present the holographic duals of the vortex loop operators in M-theory. We repeat the analysis in the string theory language in Appendix~\ref{app-D2}.

\subsection{M2-Brane Solution}
\label{sec-M2}

In this section we give the bulk description of the $1/2$ and  $1/3$ BPS vortex loop operators in the probe approximation. Since the field theory operators are supported on a curve, the object dual to them in the bulk 
must end on the boundary of $AdS_4$ along that curve. We find that the appropriate object is an array of M2-branes in the bulk.
A single M2-brane in the bulk corresponds to the case when the vortex loop operator has a non-trivial behavior only in a single $U(1)$ factor. The bulk description when the broken symmetry of the loop is $L=U(N_0)\times U(N_0)\times U(N_1)\times\cdots \times U(N_M)$ corresponds to an array of $M$ separated M2-branes.

Recall from the gauge theory analysis that for gauge group  $U(1)\times U(1)$ the conformal vortex loop operators were either $1/2$ BPS or $1/3$ BPS, while the $1/6$ BPS example had automatically enhanced supersymmetry. Indeed we find that a single M2-brane is $1/2$ BPS or $1/3$ BPS. To find $1/6$ BPS configurations, one should consider a general non-Abelian gauge group and a collection of multiple M2-branes in the bulk.

The $SU(1,1)\simeq SL(2,R)$ symmetry of the loop operators implies that the brane must span $AdS_2\subset AdS_4$. 
As explained in the previous section, the $U(1)_l\subset SO(2,3)$ symmetry that leaves the straight line or the circle invariant is    broken by the field configuration produced by a BPS loop operator. Therefore, the symmetry corresponding to shifts in the angle $\phi$ in the bulk metric   (\ref{metricmaca}) must be broken by the M2-brane embedding. Nevertheless, the $1/2$ BPS operators are invariant under a diagonal combination of $U(1)_l$ and an $U(1)_R$ symmetry, which corresponds to an  isometry of   $S^7/\bZ_k$. Therefore, the M2-brane embeddings dual the BPS loop operators wrap $AdS_2\times S^1\subset AdS_4$ and have a non-trivial profile on $S^7/\bZ_k$, which depends on the $S^1\subset AdS_4$ coordinate~$\phi$.

The $1/2$ BPS loop operators excite  a single complex scalar field $C^1$, while
the $1/3$ BPS loop operators excite  two complex scalar fields $C^1$ and $C^2$. In the bulk, we can describe both types of operators by  considering M2-branes with $w^1\neq0$, $w^2\neq0$ and $w^3=w^4=0$ in (\ref{ws}), so we set $\vartheta_2=\pi$. The relevant part of the metric on the compact manifold --- corresponding to an $S^3/\bZ_k$ --- is then given by 
\be
ds^2_{S^3/\bZ_k}=\frac{R^2}{4}\Bigg[
d\vartheta_1^2+\sin^2\vartheta_1\,d\varphi^2
+\left(\frac{2}{k}\,d\zeta+\cos\vartheta_1\,d\varphi\right)^2\Bigg]\,,
\label{lens}
\ee
where we have defined new angles
\be
\xi_1=-\frac{\varphi}{2}+\frac{\zeta}{k}\,,
\qquad
\xi_2=\frac{\varphi}{2}+\frac{\zeta}{k}\,.
\label{newangle}
\ee
Both $\zeta$ and $\varphi$ range between 0 and $2\pi$. 

We describe the M2-brane embedding  corresponding to the $1/2$ and $1/3$ BPS loop operators by
choosing the static gauge along $AdS_2\times S^1\subset AdS_4$ and considering a periodic motion on  
 $S^3/\bZ_k$
 \be
\zeta=\zeta(\phi)\,,\qquad
\varphi=\varphi(\phi)\,.
\label{zeta-varphi}
\ee
The $1/2$ BPS M2-brane embedding for the case of $k=1$ 
was found in \cite{Lunin:2007ab}, and orbifolding it gives 
both the $1/2$ BPS and the $1/3$ BPS solutions we present below. 
For completeness, we rederive the solution here. The corresponding D2-brane solution
in Type IIA string theory is described in Appendix \ref{app-D2}.

With this ansatz, the M2-brane action is given by
\be
\begin{aligned}
\cS_\text{M2}=\frac{T_\text{M2}R^3}{8}\int \Omega_{AdS_2}\,d\phi\,
&\\
&\hskip-1.3in
\left[
\cosh^2u\sqrt{\sinh^2u
+\left(\frac{2\dot\zeta}{k}+\cos\vartheta_1\,\dot\varphi\right)^2
+\dot\varphi^2\sin^2\vartheta_1}
-\cosh^3 u+1\right]\,
\label{accionmem}
\end{aligned}
\ee
with a dot representing differentiation with respect to $\phi$.  The last two terms are the contribution from the background three-form gauge
potential and $T_\text{M2}=1/4\pi^2$ is the M2-brane tension.

The equation of motion for $u$ has two solutions. The BPS solution corresponding to BPS vortex loop operators is
\be
\label{u-eom}
\cosh u=\sqrt{\sinh^2u+\left(\frac{2\dot\zeta}{k}+\dot\varphi \cos\vartheta_1\right)^2+\dot\varphi^2\sin^2\vartheta_1}\,.
\ee
The second solution is similar, with an overall factor of 2 multiplying the right-hand side. We will not discuss the other solution here.

The equation of motion for $\vartheta_1$ gives the constraint
\be
\dot\zeta\dot\varphi\sin\vartheta_1=0\,.
\ee
Seemingly there are four different solutions, with $\dot\zeta=0$, with $\dot\varphi=0$, with $\vartheta_1=0$ and $\vartheta_1=\pi$. The last three cases may, however,  be grouped together. Note that when $\sin\vartheta_1=0$, either the angle $\xi_1$ or the angle $\xi_2$ is ill defined. Therefore $\zeta$ and $\varphi$ are not independent variables \eqn{newangle}. We therefore choose in these cases to take $\varphi=0$ and end up with two cases which should be studied separately. Using \eqn{u-eom} the two cases are
\begin{align}
\label{M2-1/2}
&1.\qquad
\dot\varphi=0\,,\qquad\dot\zeta=\pm\frac{k}{2}\,,\\
\label{M2-1/3}
&2.\qquad
\dot\zeta=0\,,\qquad\dot\varphi=\pm1\,.
\end{align}

In the first case \eqn{M2-1/2} we have
 using \eqn{ws}, \eqn{newangle}
\be
w^1=\sin\frac{\vartheta_1}{2}\,e^{i\left(\pm\frac{\phi}{2}+\xi^0_1\right)}\,,\qquad
w^2=\cos\frac{\vartheta_1}{2}\,e^{i\left(\pm\frac{\phi}{2}+\xi^0_2\right)}\,,\qquad
w^3=w^4=0\,,
\label{w1w2-1/2}
\ee
where $\phi$ is the world-volume coordinate parameterizing the motion around $S^1\subset AdS_4$ and $\xi^0_1$ and $\xi^0_2$ are arbitrary constants. The choice of sign in \eqn{w1w2-1/2} corresponds to the choice   we have in making a 1/2 BPS loop operator from either a holomorphic or antiholomorphic field configuration in the gauge theory.

For this brane embedding, $w_1$ and $w_2$ are proportional to each-other and by an $SU(4)$ rotation we can go to the case with $\vartheta_1=\pi$, where $w_2=0$. This solution is dual to the $1/2$ BPS vortex with only $C^1$ turned on \eqn{singuconf}. The  supersymmetry analysis of this M2-brane embedding   is performed  in Appendix~\ref{app-M2-SUSY}, where we prove that this M2-brane is $1/2$ BPS, in agreement with the gauge theory.

In the second case \eqn{M2-1/3} we have using
\eqn{ws}, \eqn{newangle}
\be
w^1=\sin\frac{\vartheta_1}{2}\,e^{i\left(\mp\frac{\phi}{2}+\xi^0_1\right)}\,,\qquad
w^2=\cos\frac{\vartheta_1}{2}\,e^{i\left(\pm\frac{\phi}{2}+\xi^0_2\right)}\,,\qquad
w^3=w^4=0\,.
\label{w1w2-1/3}
\ee
Note that now $w_1$ and $w_2$ are not proportional to each-other, as their $\phi$ dependence has the opposite sign. This solution corresponds to the $1/3$ BPS vortex with both $C^1$ and $C^2$ turned on \eqn{singuthird}, where one field   is holomorphic and the other one antiholomorphic.
 We show in Appendix~\ref{app-M2-SUSY} that in this case the M2-brane solution is $1/3$ BPS, in agreement with the gauge theory. Furthermore we show that for $k=1,2$, where the M-theory background preserves thirty-two supercharges, this solution becomes $1/2$ BPS.

The representation of the solutions in \eqn{w1w2-1/2} and \eqn{w1w2-1/3} obscures one detail, which is the action of the $\bZ_k$ orbifold.  $w^1$ and $w^2$ are single valued complex numbers only in the universal covering space $S^7$. In that case both solutions correspond to great circles. Therefore it is also not surprising that for $k=1,2$ both solutions preserve the same number of supersymmetries. The distinction between the solutions comes when considering the orbifold, which acts also along great circles of $S^7$. The orbifold acts by shifts on the Hopf fiber. In the $1/2$ BPS case \eqn{w1w2-1/2} the two circles are completely aligned, as the M2-brane wraps the Hopf-fiber $k/2$ times. In the $1/3$ BPS case the angle between the circle that the M2-brane wraps and the circle on which the orbifold acts is $\vartheta_1$.

In all of the solutions above the phases of $w^I$ behave like $\pm\phi/2$, which is directly related to the square-root dependence in the vortex loop operator field configuration. For a single vortex we would take the M2-brane to wrap the $\phi$ circle inside $AdS_4$ once, which would require to identify $w^I\simeq -w^I$. This indeed is the case for even $k$, since the orbifold action identifies $w^I\rightarrow e^{2\pi i/k} w^I$. For odd $k$, however,  we find that there is no single M2-brane solution, as in this case the M2-brane does not close. This is the bulk realization of 
a  similar phenomenon we found on  the gauge theory side, where for odd $k$ the theory with a single vortex was ill defined.

Having established the M2-brane solutions dual to the $1/2$ BPS and $1/3$ BPS vortex loop operators we calculate now the expectation value of the  loop operators $V_C$, with $C=\bR$ or $C=S^1$, in the supergravity regime. The expectation value is determined by the on-shell action of the corresponding M2-brane
\be
\vev{V_{C}}=\exp(-\cS_\text{M2})\,.
\ee
Plugging in the classical solution into the M2-brane action 
(\ref{accionmem}) yields
\be
S_\text{M2}^\text{classical}
=\frac{T_\text{M2}R^3}{8}\int \Omega_{AdS_2}\,d\phi
=\frac{R^3}{16\pi}\int \Omega_{AdS_2}
\ee
The volume of $AdS_2$ depends on the regularization. For the straight line one 
takes the natural regularization on the Poincar\'e patch \eqn{poincare-AdS2-metric}, 
with vanishing area. For the circle one uses global $AdS_2$ \eqn{global-euclid-metric}, 
whose regularized area is $-2\pi$. For the line we find that the action vanishes 
and the expectation value of the vortex loop operator is unity.

For the circle we get
\be
S_\text{M2}^\text{classical}
=-\frac{R^3}{8}=-\frac{k}{2}\pi\sqrt{2\lambda}\,.
\ee
This is $k/2$ times the answer for a fundamental string in $AdS_4\times \CP^3$. For odd $k$, we should extend the solution to a double cover, 
so $0\leq\phi\leq4\pi$ giving $k\pi\sqrt{2\lambda}$.

This shows that the expectation value of a circular BPS loop operator at strong coupling is 
\be
\vev{V_{C}}=\exp\left[k\pi\sqrt{\lambda/2}\right]\,.
\label{final-M}
\ee

\subsection{Mapping Probe Brane and Gauge Theory Data}
\label{sec-match}

We now proceed to   identify the parameters describing the loop operators we constructed in the gauge theory with 
the parameters of the corresponding M2-branes in $AdS_4\times S^7/\bZ_k$.

The M2-brane solutions we wrote down, with a single brane winding once around the $\phi$ circle correspond to vortex loop operators where only a $1\times1$ block of the scalar fields $C^I$ is turned on. This means that in equation \eqn{levi} $N_0=N-1$ and $N_1=1$, so the unbroken gauge symmetry is $L=U(N-1)^2\times U(1)$. This case is therefore very similar to the loop operators in the $U(1)\times U(1)$ theory in Section~\ref{secabelia}, to which we now compare.

The 1/2 BPS vortex loop operator \eqn{singuconf} depends on  two parameters, a real positive number $|\beta|$ and and angular variable $\alpha$. Likewise the M2-brane solution \eqn{w1w2-1/2} (after setting $\vartheta_1=\pi$ by an $SU(4)$ rotation) depends on  two parameters: $u$, which determines the radius of curvature of the $AdS_2\times S^1$ worldvolume metric and an angular variable $\xi_1^0$, which gives the relative phase  between the circle in $AdS_4$ and the Hopf fiber in $S^7/\bZ_k$. We propose to identify
\be
\sinh u={1\over \pi\sqrt{2\lambda}}\,|\beta|\,,
\qquad \qquad
\xi_1^0=\frac{2\pi\alpha}{k}\,.
\label{identifico}
\ee
This mapping of parameters is determined by the symmetries that the solutions preserve and those they break, up to constants, which are guessed from the analogy with the surface operators in $\cN=4$ SYM \cite{Drukker:2008wr}.

The 1/3 BPS vortex loop operator in the Abelian theory \eqn{singuthird} depends on four real parameters: A pair of complex numbers $(\beta_1,\beta_2)$ subject to the identification 
$(\beta_1,\beta_2)\simeq e^{i\theta}(\beta_1,\beta_2)$ and an angular variable $\alpha$. The associated M2-brane \eqn{w1w2-1/3} depends,  as in the $1/2$ BPS case, on $u$, but now the solution depends also on  the angle $\vartheta_1$ measuring the angle between the circle that the M2 wraps and the circle on which the $\bZ_k$ orbifold acts. In addition,  the brane embedding depends on  two phases $\xi_1^0$ and $\xi_2^0$,  which can be also rearranged, as in \eqn{newangle} into
\be
\varphi_0=\xi_2^0-\xi_1^0\,,\qquad
\zeta_0=\frac{k}{2}\left(\xi_1^0+\xi_2^0\right).
\label{zeta0phi0}
\ee

To find the map between the gauge theory parameters and the parameters of our
M2-brane solution \eqn{M2-1/3}, we recall that the two homogeneous coordinates $w^1$ and $w^2$ defined in \eqn{ws}  correspond to the fields $C^1$ and $C^2$ in the gauge
theory. The vortex singularity has the following form \eqn{singuthird}
\be
C^1=\frac{\beta_1}{\sqrt{z}}\,,\qquad
C^2=\frac{\beta_2}{\sqrt{\bar z}}\,.
\ee
Using the map $C^1\to w^1$ and $C^2\to w^2$, we find that
\be
\tan\frac{\vartheta_1}{2}\,e^{-i\varphi}
=\frac{w^1}{w^2}
= \frac{C^1}{C^2}
=\frac{\beta_1}{\beta_2}\,\sqrt{\frac{\bar z}{z}}
=\frac{\beta_1}{\beta_2}e^{-i\phi}\,.
\ee
Comparing with the M2-brane solution we see that this loop operator corresponds to the choice of positive sign in equation \eqn{M2-1/3}, the choice of negative sign corresponding to a vortex loop operator where the role of holomorphic fields is replaced by antiholomorphic fields. Using \eqn{w1w2-1/3} and \eqn{zeta0phi0} and the fact that on the solution $\varphi=\phi+\varphi_0$, we find that 
\be
\tan\frac{\vartheta_1}{2}\,e^{-i\varphi_0}={\beta_1\over \beta_2}\,.
\ee
The remaining two parameters on the M-theory side are identified in a similar way to the $1/2$ BPS case. Explicitly, the proposed identification of parameters of the 1/3 BPS loop operator and of the 1/3 BPS M2-brane is given by
\be
\begin{aligned}
\sinh u&=\frac{1}{\pi\sqrt{2\lambda}}\sqrt{|\beta_1|^2+|\beta_2|^2}\,,
\qquad&
\tan\frac{\vartheta_1}{2}&=\left|\frac{\beta_1}{\beta_2}\right|,\\
\varphi_0&=\arg\frac{\beta_2}{\beta_1}\,,
\qquad&
\zeta_0&=2\pi\alpha\,.
\end{aligned}
\label{map}
\ee

The $1/2$ BPS case is recovered by taking $\vartheta_1\to\pi$. As we saw, our choice of holomorphic fields corresponds to the choice of positive sign  in \eqn{M2-1/3}. Because $\zeta$ and $\varphi$ appear with opposite signs in \eqn{newangle}, we conclude that in the $1/2$ BPS case we should take the negative sign in \eqn{M2-1/2} to match with the holomorphic  vortex loops in the gauge theory. The positive sign corresponds to antiholomorphic vortex loop operators.

Turning to the non-Abelian case, all the BPS vortex loop operators constructed in Section~\ref{sec-nonabelian} are described by block-diagonal matrices, where in each block there is a copy of a $1/3$ BPS vortex of the Abelian theory (possibly rotated). This is mirrored in the M-theory dual, where each block in the matrix should be represented by a single M2-brane.

Specifically, the general BPS vortex in the non-Abelian theory depends on $M$ integers $N_1,\cdots N_M$  and has an unbroken gauge symmetry $U(N_0)^2\times U(N_1)\times \cdots\times U(N_M)$, where $N_0=N-\sum_{l=1}^M N_l$. The natural identification is to represent in M-theory each block by a single M2-brane wrapped $N_l$ times around the $\phi$ circle in $AdS_4$. The rest of the data in the classification of the vortex loop $V_C$ are the collection of numbers $(\beta_I^{(l)}, \alpha^{(l)})$. They are related to the M2-brane parameters as in \eqn{map}, with the only extra new information being that in the $1/6$ BPS case \eqn{holomorphic6}, \eqn{holomorphic6b} in each block the vortex may have different ratios of $\beta_1^{(l)}$ and $\beta_2^{(l)}$ and of $\beta_3^{(l)}$ and $\beta_4^{(l)}$, which translates in an obvious way to a choice of $S^3/\bZ_k$ in which the M2-brane is embedded.

As noted before, for odd $k$ the single M2-brane configuration is inconsistent, as it does not close onto itself. This is the M-theory manifestation of the fact that some operators in the gauge theory are not single valued in the presence of vortex loop operators for odd $k$. This problem is avoided, though, when all M2-branes are wrapped an even number of times around the $\phi$ circle. According to the preceding prescription, this happens when the integers $N_l$ with $l=1,\cdots M$ parameterizing the unbroken gauge group are all even. Indeed we saw also on the gauge theory side that to construct consistent vortices at odd $k$ requires all $N_l$ to be even and that it involves a non-Abelian twist \eqn{holomorphic/2} and \eqn{twist}.

\subsection{Correlator with Local Operators}
\label{sec-local-M}

We want to calculate, using the preceding probe M2-brane description,
the correlator of a vortex loop operator $V_C$ with a chiral primary operator. This 
is the bulk M-theory
analog of the calculation performed in the gauge theory in Section~\ref{sec-local-CS}. 
We will perform this computation for the $1/2$ BPS solution \eqn{M2-1/2}. The necessary harmonic analysis on $AdS_4\times S^7/\bZ_k$ and the analysis of supergravity fluctuations needed for this computation are detailed in 
Appendices~\ref{app-harmonics} and~\ref{app-SUGRA-modes}, 
based on 
\cite{Biran:1983iy,Castellani:1984vv,Bastianelli:1999bm}. Similar
calculations have been performed in the context of $AdS_7\times S^4$ in
\cite{Corrado:1999pi,Chen:2007zzr} and in the context of $AdS_5 \times
S^5$ in 
\cite{Berenstein:1998ij,Semenoff:2004qr,Giombi:2006de,Semenoff:2006am,Drukker:2008wr}.

A chiral primary operator in ${\cal N}=6$ Chern-Simons theory $\cO^A$ corresponds in the dual supergravity description to 
a four dimensional scalar field $s^A$ propagating in $AdS_4$. The   correlator of a vortex loop operator $V_C$ and a chiral primary operator is determined by the normalizable mode of $s^A$ produced by   the probe M2-brane. Therefore, we must first compute the linearized coupling of the M2-brane to the supergravity field $s^A$. 
This is found by varying the membrane
action with respect to the spacetime metric and three-form field
\be
\delta S_\text{M2}
=\frac{T_\text{M2}R^3}{8}\int d^3\sigma\left[
\frac{1}{2}\,\sqrt{\det g_{ab}}\,g^{ab}\, \p_a X^M \p_b X^N\,h_{M N}-P[\d C_3]\right]\,.
\label{fluctua}
\ee
Here $g_{ab}$ is the induced metric on the brane, which is that of $AdS_2\times S^1$ 
with radius $\cosh u$. $h_{MN}$ and $\d C_3$ are the  fluctuations of the metric and three-form field and the indices $M$ and $N$ go over all eleven dimensions.

Since we are interested in the correlator with a chiral primary operator, which is dual to the bulk field $s^A$, we must relate the fluctuations
of the metric and three-form  in (\ref{fluctua}) to $s^A$. To linear order, the harmonic expansion of  the metric and three-form fluctuations are given by
\eqn{SUGfluc}, \eqn{SUGmodes}
\bsp 
h^A_{\m\n}& = \frac{4}{(J+2)}\left[ \nabla_\m \nabla_\n
  +\frac{J(J+6)}{8} g_{\m\n} \right] s^A - \frac{7J}{6} g_{\m\n} s^A\\
h^A_{\a\b} &= \frac{J}{3}\, g_{\a\b}\, s^A,\\
\delta C^A_{\m\n\r} &= 2 \, \varepsilon_{\m\n\r\l}\, \nabla^\l s^A \,,
\end{split}
\label{relacional}
\ee
where $\mu,\nu,\cdots$ are indices along $AdS_4$ and $\alpha,\beta,\cdots$ are indices along $S^7/\bZ_k$. The integer $J$ determines the eigenvalue of the Laplacian on $S^7/\bZ_k$ of the corresponding spherical harmonic  and is equal to twice the conformal dimension $\Delta$ of the dual operator (see Appendix \ref{app-harmonics}).

Using the coordinate system \eqn{metricmaca}, \eqn{poincare-AdS2-metric}, and the metric in \eqn{lens}, the relevant part of the bulk metric is given by
\be
ds^2 = \frac{\cosh^2u}{z^2}\left( dt^2 + dz^2 \right) +du^2 + \sinh^2u \,d\phi^2+ds^2_{S^3/\bZ_k}\,.
\ee
On the 1/2 BPS solution \eqn{M2-1/2}, where the worldvolume coordinates are $t'$, $z'$ and $\phi'$ and where $\vartheta_1=\pi$ and $\dot\zeta =k/2$,
  we find that the induced metric on the M2-brane is given by
\be
g^{ab}  = 
\frac{1}{\cosh^2 u}\text{diag} \left(z^2 ,\,z^2,\, 1 \right)\,.
\ee
The various fluctuations   appearing in \eqn{fluctua} are given by
\be
\begin{aligned}
\p_a X^M \p_b X^N\,h^A_{MN} 
&= \text{diag} \left( h^A_{tt},\, h^A_{zz}, \, h^A_{\phi \phi}  + \frac{J}{3}s^A \right)
\\
P[\d C_3^A] &= 2 \,\frac{\cosh^2u \,\sinh u}{z^2}\, \p_u s^A\,.
\end{aligned}
\ee
We then find that the linearized coupling of the bulk field $s^A$ to the M2-brane worldvolume is given by 
\begin{align}
\delta S_\text{M2}
=\frac{T_\text{M2}R^3}{8}\int &dz\,dt\,d\phi
\frac{\cosh u}{2(J+2)} \Biggl[ 
4\left(\p_z^2 +\p_t^2 +\frac{1}{z^2} \p_\phi^2 
-\frac{J-1}{z^2}\cosh u\sinh u\,\p_u \right)\nonumber\\
&\hskip2cm
-\frac{2J(J-1)}{3} \frac{(3\cosh^2u-1)}{z^2} + \frac{J(J+2)}{3z^2} \Biggr] s^A Y^A\,,
\end{align}
where $Y^A$ are $S^7/\bZ_k$ spherical harmonics.

We now consider the insertion of the local chiral primary operator corresponding to $s^A$ 
at the $AdS_2\times S^1$ boundary point labeled by $(t,z,\phi)$. The expression for $s^A$ at a point $(t',z',\phi',u)$ along the brane once 
a source $s_0^A(t,z,\phi)$ is specified on the boundary is given by integrating the 
bulk-to-boundary propagator from the point at the boundary to a point in the brane. The bulk-to-boundary propagator in 
our coordinate system is given by\footnote{We recall that $\Delta=J/2$.}
\be
G(u,z',t',\phi')= c_J \frac{z'^\D}{\cosh^\D u\, D^\D} \,,
\label{prop}
\ee
where
\be
D \equiv (t'-t)^2 +z'^2 + z^2 -2 z' z \tanh u \, \cos(\phi' - \phi)\,,
\ee
and $c_J$ is a normalization constant given by (\ref{cJ}), which guarantees   that the bulk computation of the two-point function of the corresponding chiral primary operator is unit normalized as in (\ref{unidad}). Acting with the derivatives on the propagator and 
simplifying we find that the correlator of the 
vortex loop operator with the chiral primary operator dual to $s^A$ is
\be
\begin{aligned}
\frac{\vev{V_C\cdot \cO^A}}{\vev{V_C}}
&=-\frac{T_{M2}R^3}{8}  \int_{-\infty}^\infty dt' \,
\int_0^\infty dz'\, \int_0^{2\pi} d\phi'\, 
\frac{2\Delta\,c_J}{\cosh^{\Delta+1}u}\frac{2z'^{\Delta}
  z^2}{D^{\Delta+2}} \,Y^A\\
& = -\frac{T_{M2}R^3}{8} \frac{4c_J \Delta\sqrt{\pi}\,z^2}{\cosh^{\Delta+1}u}
\frac{\Gamma(\Delta+3/2)}{(\Delta+1)!} 
\int_0^{2\pi} d\phi'\,Y^A
\int_0^\infty dz'\,\frac{z'^\Delta}{\hat D^{\Delta+3/2}}\,,
\end{aligned}
\ee
where $\hat D \equiv z'^2 + z^2 -2 z' z \tanh u \, \cos(\phi' - \phi)$. 

The $z'$ integration yields (first scaling $z$ out)
\bsp 
\int_0^\infty dz'\, \frac{z'^{\Delta}}{\hat D^{\Delta+3/2}} &=
\frac{1}{z^{\Delta+2}} \int_0^\infty dz' \,
\frac{z'^{\Delta}}{(1+z'^2-2z'\tanh u \cos \hat\phi)^{\Delta+3/2}}\\ 
&=\frac{1}{z^{\Delta+2}} \frac{\sqrt\pi}{2^{\Delta+1}}\, 
\frac{\Delta!}{\G(\Delta+3/2)}\, \frac{1}{(1-\tanh u\, \cos
  \hat \phi)^{\Delta+1}},
\end{split}
\ee
where $\hat \phi = \phi' - \phi$.

Lastly we perform the $\phi'$ integration. Here we need the explicit form of the 
spherical harmonics from Appendix~\ref{app-harmonics}. For the $1/2$ BPS 
M2-brane solution we have $\vartheta_1=\pi$ and $\xi_1=\xi_1^0-\phi'/2$. 
Using that $P_n^{\alpha,\beta}(-1)=(-1)^n{n+\beta\choose n}$ we get from 
\eqn{normalized}
\be
Y_{\Delta,p}(\vartheta_1=\pi)
=(-1)^{\Delta_-}\sqrt{\frac{(\Delta_++2)!\,(\Delta_-+2)!}{2(2\Delta+2)!}}\,
e^{ipk(\xi_1^0-\phi'/2)}\,,
\ee
where $\Delta_\pm=\Delta\pm \frac{pk}{2}$. Therefore the last integral is of the form
\be
\frac{1}{\cosh^{\Delta+1}u}
\int_0^{2\pi}d\hat\phi\, \frac{e^{ipk\hat\phi/2}}{(1-\tanh u\, \cos \hat \phi)^{\Delta+1}} 
= 2\pi\,\frac{\Delta_-!}{\Delta!}\,P_{\Delta}^{{pk/2}}(\cosh u),
\ee
where $P^m_n(x)$ is an associated Legendre function.

Assembling everything together, we find
\begin{align}
\frac{\vev{V_C\cdot \cO_{\Delta,p}}}{\vev{V_C}}
=&\, \frac{(-1)^{\Delta_-+1}}{(2\pi^2\lambda)^{1/4}}\frac{\pi}{2\sqrt2}
\, \D_-! \,
\sqrt{\frac{(\Delta_++1)(\Delta_++2)(\Delta_-+1)(\Delta_-+2)}
{(2\Delta)!\,(\Delta+1)}}
\nonumber\\&\hskip5cm\times
\frac{e^{ipk(\xi_1^0-\phi/2)}}{z^\Delta}P_{\Delta}^{{pk/2}}(\cosh u)\,,
\label{M2-correlator}
\end{align}
where $\cO_{\Delta,p}$ are the $SU(3)$ invariant chiral primary operators.
For the first few values of $\D$, we find
\bsp
\frac{\vev{V_C\cdot \cO_{1,0}}}{\vev{V_C}}
&= \frac{1}{(2\pi^2 \l)^{1/4}} \frac{3\sqrt{2}\,\pi}{4}\frac{\cosh u}{z}\,,\\
\frac{\vev{V_C\cdot \cO_{2,0}}}{\vev{V_C}}
&= -\frac{1}{(2\pi^2 \l)^{1/4}} \frac{\pi}{2}\frac{3\cosh^2 u-1}{z^2} \,,\\
\frac{\vev{V_C\cdot \cO_{3,0}}}{\vev{V_C}}
&= \frac{1}{(2\pi^2 \l)^{1/4}} \frac{\sqrt{10}\,\pi}{8}\,
\frac{5\cosh^3 u - 3\cosh u}{z^3} \,.
\end{split}
\ee
Using that the leading power in the Legendre polynomials is 
$P_n^0(x)={2n\choose n}\frac{x^n}{2^n}+\cdots$ we can write the leading 
term for operators with arbitrary $\Delta$ as
\be
\frac{\vev{V_C\cdot \cO_{\Delta,0}}}{\vev{V_C}}
=\frac{1}{(2\pi^2 \l)^{1/4}}\frac{\pi(\Delta+2)}{2\sqrt2}
\sqrt{\frac{(2\Delta-1)!!\,(\Delta+1)}{(2\Delta)!!}}\,\cosh^\Delta u
+\cdots
\ee

Also using $P_n^n(x)=(-1)^n(2n-1)!!(1-x^2)^{n/2}$ we have for the case of 
$pk=2\Delta$ that
\be
\frac{\vev{V_C\cdot \cO_{\frac{pk}{2},p}}}{\vev{V_C}}
=-\frac{(-i)^\Delta}{(2\pi^2\lambda)^{1/4}}\frac{\pi}{\sqrt2}
\sqrt{\frac{(2\D+1)!!}{(2\D)!!}}
\left(\frac{e^{i(2\xi_1^0-\phi)}}{z}\right)^\Delta\sinh^\Delta u\,.
\ee

\subsubsection*{$\bullet\quad $Comparison to Semiclassical Calculation}

It is interesting to try to compare our results here to those we found by 
semiclassical techniques in $\cN=6$ Chern-Simons theory in Section~\ref{sec-local-CS}. 
Like the expectation value of the vortex loop operator, the correlation 
function of a loop operator with a chiral primary operator also gets non-trivial 
quantum corrections to all orders in the 't Hooft coupling. This is in contrast to 
the analog computation with  surface operators in $\cN=4$ SYM, whose correlators 
with local operators seem to get   quantum corrections only to a finite loop order
\cite{Drukker:2008wr}.

Using \eqn{map} we can represent the result of our calculation performed at strong coupling \eqn{M2-correlator}
in terms of the gauge theory variables
\be
\frac{\vev{V_C\cdot \cO_{\Delta,p}}}{\vev{V_C}}
\sim\frac{1}{(2\pi^2\lambda)^{1/4}}
\frac{e^{4\pi ip\alpha}}{\sqrt{z^{\Delta_+} \bar z^{\Delta_-}}}
P_{\Delta}^{{pk/2}}\left(\sqrt{1+\frac{|\beta|^2}{2\pi^2\lambda}}\,\right).
\ee
We omitted all numerical factors in this expression. Also we replaced $ze^{i\phi}\to z$, 
which is the holomorphic coordinate in the plane transverse to the loop that we used 
in the gauge theory calculation.

There are some general features we would like to point out in this expression. First, the 
dependence on the holomorphic and anti-holomorphic coordinates $z$ and $\bar z$ is 
as would be expected for a field of dimension $\Delta$ and $U(1)_B$ charge $pk$. A 
feature that might seem surprising at first is the appearance of $\alpha$, the holonomy of the gauge 
field, in the correlator of a scalar operator. This happens only in the case of non-zero 
$p$, when the chiral primary operator is not made purely of scalar fields, but also 
carries a monopole charge, and hence the dependence also on the holonomy of the gauge field.

In the gauge theory calculation in the semiclassical approximation we found that for $p=0$ \eqn{pert-correlator}
\be
\frac{\vev{V_C\cdot {\cal O}_{\Delta,0}}}{\vev{ V_C}}
\sim\frac{1}{|z|^\Delta}
\left(\frac{4\pi}{\lambda}\right)^\Delta
|\beta|^{2\Delta}\,.
\ee
This semiclassical result will receive quantum corrections. A simple class of
quantum corrections involves  self-contractions 
of the scalar fields in the operator $\cO_{\Delta,0}$. Since 
a pair of scalars  $CC^\dagger$ have to be contracted with another pair,
this class of graphs give quantum corrections in $\lambda^2/|\beta|^4$ only to a finite loop order, $\Delta_-$.

The numerical coefficients appearing in the correlators seem to get renormalized between weak and strong 
coupling as well as  the form of the expansions  in $\beta$ and $\lambda$, where
\be
\frac{|\beta|^2}{\lambda}\to \frac{|\beta|}{\sqrt{\lambda}}\,,
\ee
which is a generalization of the scaling of the coupling from weak to 
strong coupling that appears in other calculations in this theory. In addition we 
find that the correlators in M-theory get an extra factor of $\lambda^{-1/4}$. 
The only thing that seems to match is the fact that the calculation involves 
also a polynomial of degree $\Delta_-$ in the respective expansion parameters at
weak coupling and strong coupling.

\subsection{``Bubbling'' M-Theory Geometries}
\label{sec-bubbles}

In Section ~\ref{sec-M2} we found   
 the description of a vortex loop operator in terms of probe branes  embedded in 
$AdS_4\times S^7/\bZ_k$. The probe brane description of a vortex loop operator   is valid 
as long as the number of M2-branes is much smaller than $N$. The operators for which 
the probe approximation is valid have  $N_0\sim N$ in \eqn{levi}. When the number
of branes describing the operator  is of order $N$, the gravitational backreaction of the M2-branes cannot be neglected, 
and the proper dual description of the operator is in terms of ``bubbling geometries'' 
\cite{Lin:2004nb}.

The supergravity solutions capturing  the backreaction of the $1/2$ BPS M2-brane solutions of Section~\ref{sec-M2} can be written down by a simple modification of a class of bubbling supergravity  solutions found by Lunin in \cite{Lunin:2007ab}.
The supergravity solutions constructed by Lunin  posses an $SL(2,R)\times SO(6)$ symmetry and can be obtained by a double Wick-rotation of the 
   bubbling  solutions describing  giant gravitons in $AdS_4\times S^7$ 
\cite{Lin:2004nb}. By appropriately orbifolding,   we   find solutions where the symmetry is generically broken to 
 $SL(2,R)\times SU(3)$.

The metric ansatz studied in \cite{Lunin:2007ab} (before orbifolding) has factors of $AdS_2$ and $S^5$ which make  explicit the desired $SL(2,R)\times SO(6)$ symmetries. 
The metrics can be written as
\be
\begin{aligned}
ds^2=&\,e^{2\omega}(y^2e^{-6\omega}-1)(d\chi+V_idx^i)^2
+\frac{e^{-4\omega}}{4(y^2e^{-6\omega}-1)}(dy^2+e^D(dx_1^2+dx_2^2))
\\&
+e^{2\omega}ds_{S^5}^2+\frac{1}{4}y^2e^{-4\omega}ds_{AdS_2}^2\,.
\end{aligned}
\label{metric-bubbles}
\ee
The  supergravity solutions   are completely determined by a function $D$, which satisfies a 3-dimensional Toda equation in 
the coordinates $x_1$, $x_2$ and $y$. 
The warp factor $\omega$ and the vector field $V_i$ are given in terms of $D$ by
\be
e^{-6\omega}=\frac{\partial_yD}{y(1-y\partial_y D)}\,,\qquad
V_i=\frac{1}{2}\,\varepsilon_{ij}\partial_iD\,.
\ee
These solutions    also have a four-form field strength turned on. It is given by ($\star_3$ is the 
Hodge-duality operator on the base manifold parametrized by $x_1$, $x_2$ and $y$).
\be
F_4=
\left(d\left[-4y^3e^{-6\omega}(d\chi+V)\right]
+2\star_3\left[e^{-D}y^2\left(\partial_y\frac{1}{y}\partial_y e^D\right)
+y\partial_i\partial_y D dx^i\right]\right)\wedge \Omega_{AdS_2}\,.
\ee

As already mentioned, the solution is completely determined by the function $D$, which  solves the equation
\be
(\partial_1^2+\partial_2^2)D+\partial_y^2e^D=0\,.
\ee
One needs to analyze this equation and the allowed boundary conditions and singularities that 
give rise to smooth geometries.

In \cite{Lunin:2007ab} two classes of solutions were considered, of which the second class is the  relevant one  for us. For these solutions the $y$ coordinate extends from 0 to infinity. At $y=0$ the function $D$ develops a singularity, $D\sim\log y$, where the radius of the $S^5$ shrinks to zero size, but the full metric remains regular.

The other allowed singularities for the function $D$ occur along semi-infinte rays extended in the $y$ direction, with $y\geq y^{(l)}$ at fixed $x_i^{(l)}$. Near the rays  $D\sim-\log|x-x^{(l)}|$ and at  the tip of each ray the circle parametrized by $\chi$  in \eqn{metric-bubbles} shrinks to zero size, but again in a regular fashion.%
\footnote{There is an alternative description of these solutions where the rays are replaced by finite rods with $0\leq y\leq y^{(l)}$, but the mapping between it and the probe brane picture is more complicated. Note, though, that in our description the $x$ plane is double valued.}
A bubbling supergravity solution is completely determined by specifying the ray structure, which is characterized by the position of the ends of the rays $(x_i^{(l)},y^{(l)})$, for $l=0,\ldots, M$.

To adapt these solutions to the problem at hand,  we need to perform a $\bZ_k$ orbifold of some circle in the ten dimensional geometry. The geometries in \eqn{metric-bubbles} have a $U(1)$ isometry, which acts by shifts on the   coordinate $\chi$ spanning the circle. In addition to the manifest circle, there is the $S^5$ which can be written as a circle fibration over $\CP^2$. If we set $w^1=0$ in \eqn{ws} we can write the $S^5$ metric in the form
\be
ds^2_{S^5}=ds_{\CP^2}^2+(d\zeta'+\tilde\omega)^2\,,\qquad
\zeta'=\frac{\xi_2+\xi_3+\xi_4}{3}\,.
\ee
We take the $\bZ_k$ orbifold to act on the angle
\be
\zeta=\frac{\chi+3\zeta'}{4}\,.
\label{bubble-zeta}
 \ee
As we show below, for the $AdS_4\times S^7$ solution indeed $\zeta=(\xi_1+\xi_2+\xi_3+\xi_4)/4$, which is the desired orbifold direction.

This orbifold action is singular at any point in the geometry where the $\zeta$ circle shrinks to zero size. Since there are no mixed metric components for the coordinates $\chi$ and $\zeta'$, this   happens only at the locus where  both circles shrink to zero size. The $\zeta'$ circle shrinks when the radius of $S^5$ goes to zero, which as we reviewed occurs  at $y=0$. The $\chi$ circle, on the other hand, shrinks to zero at the tip of each of the rays at $(x_i,y)=(x_i^{(l)},y^{(l)})$. Since regular solutions have $y^{(l)}>0$, these two conditions never coincide, and consequently the orbifold action has no fixed points. Therefore, we can orbifold the solutions in \cite{Lunin:2007ab}  and obtain completely regular backgrounds.

In order to understand the relation between the orbifolded bubbling geometries and probe M2-branes it is illuminating to describe the  $AdS_4\times S^7$ solution in this form. This solution corresponds to a single ray located at $x_i^{(0)}=0$ with $y^{(0)}=R^3$. 
It can be expressed as \cite{Lunin:2007ab}
\be
\begin{gathered}
x_1+ix_2=x\,e^{i\psi}\,,\qquad
x=R^3\sinh u\sin^2\frac{\vartheta_1}{2}\,,\qquad
y=R^3\cosh u\cos^2\frac{\vartheta_1}{2}\,,\\
e^{2\omega}=R^2\cos^2\frac{\vartheta_1}{2}\,,\qquad
e^D=\cot^2\frac{\vartheta_1}{2}\,,\qquad
V=-\frac{\sinh^2u}{2(\sinh^2u+\sin^2\frac{\vartheta_1}{2})}\,d\psi\,.
\end{gathered}
\label{bubble}
\ee
This completely matches the metrics \eqn{metricmaca}, \eqn{global-euclid-metric}, \eqn{S7-metric} once the following  identification of  angles is made
\be
\chi=\xi_1\,,\qquad
\psi-2\chi=\phi\,.
\label{bubble-angles}
\ee
The remaining angles $\vartheta_2$, $\vartheta_3$, $\xi_2$, $\xi_3$ and $\xi_4$ 
parametrize $S^5$. Orbifolding the $\zeta$ circle \eqn{bubble-zeta} indeed gives 
the metric on $AdS_4\times S^7/\bZ_k$.

It is easy to identify the probe brane solution \eqn{M2-1/2} of Section~\ref{sec-M2} in this construction. It corresponds to two   rays, one at $x_i^{(0)}=0$ and $y\geq y^{(0)}=R^3$, which generates the $AdS_4\times S^7/\bZ_k$ geometry \eqn{bubble} and another at a point $x_i^{(1)}$ with very small $y^{(1)}$. This second singularity represents an M2-brane at $\vartheta_1=\pi$, $\sinh u=|x^{(1)}|/R^3$ and $\xi_1^0=\psi^{(1)}/2$. The number of coincident M2-branes (or their wrapping number) is related to the value of $y^{(1)}$ by a rather complicated integral given in \cite{Lunin:2007ab}.

Before orbifolding, the geometries found in  \cite{Lunin:2007ab} preserve sixteen supercharges. We expect that orbifolding the $\zeta$ circle by $\zeta\simeq\zeta+2\pi/k$ will break the supersymmetry down to twelve supercharges. These orbifolded geometries     provide the  dual gravitational description of the $1/2$ BPS vortex loop operators $V_C$ in the supergravity  regime, when the probe approximation breaks down,  corresponding to the case when the number of probe M2-branes is of order $N$. The parameters of the solutions indeed match those of the $1/2$ BPS vortex loop operators. The vortex loop operator with   gauge group broken down to $L=U(N_0)^2\times U(N_1)\times\cdots U(N_M)$ gets  identified with the bubbling geometry consisting of   $M+1$ rays. One of the rays is at $x_i^{(0)}=0$ while the remaining  $M$ others are at positions%
\footnote{we use \eqn{identifico} and $R^3/4k=\pi\sqrt{2\lambda}$.}
\be
x_1^{(l)}+ix_2^{(l)}=4k|\beta^{(l)}|e^{4\pi i\alpha^{(l)}}\,,
\ee
where $\alpha^{(l)}$ and $\beta^{(l)}$ are the parameters characterizing the vortex loop operator \eqn{holomorphic}, \eqn{holomorphicgauge}. The integers $N_l$ correspond to the length of the rays $y^{(l)}$.

 The ``bubbling'' geometry has weak curvature everywhere when $\lambda$ is large and all  rays  are well seperated and the values of $y^{(l)}$ are all comparable (as mentioned above $y^{(l)}\sim0$ corresponds to a probe brane). In this regime eleven dimensional supergravity on this background provides the most reliable description of the vortex loop operators we have constructed in this paper.

As discussed earlier, some of the $1/3$ BPS vortex loop operators have the special property that they become $1/2$ BPS   for $k=1,2$. In the probe approximation all the M2-branes wrap the same great circle on the covering space $S^7$, but this circle is not aligned with the direction of the orbifold. In these cases the $1/3$ BPS vortices can also be described by an orbifold of the $1/2$ BPS geometries above, by letting $\bZ_k$ act on a different angle than $\zeta$ \eqn{bubble-zeta}. It would be interesting to understand the details of this as well as to find the most general geometry preserving eight supercharges, and representing the most general $1/3$ BPS vortex loop operator.

It is possible to calculate the correlation function of the vortex loop operators with chiral primary local operators, as we did in the gauge theory in Section~\ref{sec-local-CS} and in the probe approximation in Section~\ref{sec-local-M}, also in the bubbling geometry description. This was carried out for the case of surface operators in $\cN=4$ SYM in \cite{Drukker:2008wr}  using  techniques from \cite{Skenderis:2007yb}. It would be interesting to work out the details of the formalism in this case too and have another set of results to compare with the gauge theory \eqn{pert-correlator} and with the probe \eqn{M2-correlator} calculations.

\section{Discussion and Summary}
\label{sec-discussion}

The ${\cN=6}$ supersymmetric Chern-Simons theory of Aharony, Bergman, Jafferis and Maldacena \cite{Aharony:2008ug} provides a concrete duality between a three dimensional interacting conformal field theory and quantum gravity on spaces with $AdS_4\times S^7/\bZ_k$ asymptotics. The gravitational description of three dimensional field theories provides us with new tools to study the behaviour of these theories  at strong coupling, which may lead to new insights on the behaviour  of strongly coupled three dimensional theories describing various physical systems.

In this paper we have constructed novel disorder operators in Chern-Simons-matter theories. These operators, apart from providing a new tool to study holography, may find applications in other Chern-Simons-matter theories, known to describe some physical systems. In particular, these operators have a singularity along a curve in spacetime for the matter fields and gauge fields in the theory. These are codimension two vortex field configurations, 
 not unlike the vortices in superconductors or other physical systems. These operators may serve as order parameters for new phases in these theories.

The codimension two singularities characterizing these loop operators in Chern-Simons-matter theories are similar to the codimension two singularities describing surface operators in ${\cal N}=4$ SYM  \cite{Gukov:2006jk} (see also \cite{Gomis:2007fi}). Recently \cite{Drukker:2008wr}, these surface operators have been used to perform precision calculations across the different coupling regimes: Weakly coupled semiclassical gauge theory, D-branes in $AdS_5\times S^5$ and ``bubbling'' supergravity solutions. For all the calculations in that theory there seems to be remarkable agreement between the various regimes. For the most detailed calculation, the correlator between a surface operator and a chiral primary operator, the supergravity result can be rewritten in the gauge theory language to yield the precise semiclassical answer plus a finite series of quantum corrections, providing strong evidence that these operators only receive a small subset of the possible quantum corrections. Similar calculations across the various different regimes of coupling have been performed in \cite{Semenoff:2001xp, Okuyama:2006jc, Giombi:2006de,Gomis:2008qa} for Wilson loops in ${\cal N}=4$ SYM.

While the calculations performed in this paper are indeed similar to those in \cite{Drukker:2008wr}, just like in other corners of the $AdS_4$/CFT$_3$ duality, the agreement is not as clean as in the case of $AdS_5$/CFT$_4$. Yet, since the agreement in the case of surface operators in the four dimensional CFT is so clean, we hope that understanding vortex loop operators in the three dimensional CFT will help us learn how to perform precision calculations in the $AdS_4$/CFT$_3$ duality.

We gave a rather detailed exposition on disorder vortex loop operators --- both from the gauge theory point of view and from M-theory --- which we hope will be a useful starting point for a more detailed study of these objects. But for the benefit of the casual reader we provide now a summary of our results organized in a different way than the main text --- intertwining results from the gauge theory and M-theory pictures.

The most symmetric object we have described is the $1/2$ BPS vortex loop operator. It turns on only one of the four complex scalar fields and to preserve conformal symmetry it has a singularity along a line or a circle in space-time \eqn{singuconf}. In the M-theory dual it is described by an M2-brane occupying a hypersurface $AdS_2\times S^1\subset AdS_4$. Going around the $S^1$, the brane also wraps $k/2$ times the orbifolded circle on $S^7/\bZ_k$. Consequently, a single abelian vortex loop is well defined only for the theory with even $k$. At odd $k$ one needs to compensate for this by ``doubling'' the vortex, so in the M-theory picture it wraps the orbifold circle $k$ times. We have pointed out throughout the text the subtleties that arise when trying to define the vortex loops at odd $k$ and explained how they are resolved.

This vortex loop preserves twelve out of the twenty-four supercharges of the theory. In the case of the line, where it is a holomorphic function in the transverse plane, it preserves six super-Poincar\'e generators and six superconformal ones (for the circle it is linear combinations of both). In fact, the only other known operators preserving six Poincar\'e supercharges are the ``baryonic'' local operators $(C)^{pk}$. Other chiral primary local operators preserve only four of the Poincar\'e supercharges (of course, all chiral primaries preserve also all the super-conformal generators, a property not shared by non-local operators).

The $1/2$ BPS vortices have a close cousin which is $1/3$ BPS. On the gauge theory side it corresponds to turning on a second scalar field and giving it an anti-holomorphic dependence in the transverse space. In M-theory it is described by a similar M2-brane occupying the same hypersurface inside $AdS_4$, only that now the motion on $S^7/\bZ_k$ is on another circle, at arbitrary angle with respect to the direction of the orbifold.

These vortex loops preserve eight of the twenty-four supercharges of the vacuum, and in the case of the line four of the twelve super-Poincar\'e generators. In fact, there is a close analogy to the spectrum of chiral primary operators, where after the orbifold projection some of the $1/2$ BPS operators retain six super-Poincar\'e generators while all the others retain only four. The most symmetric ones, $(C)^{pk}$ are the ones whose momentum is aligned with the orbifold direction. Likewise the same M2-brane solution on $S^7$ is the dual of the $1/2$ BPS vortex loop and the $1/3$ BPS vortex loop, depending on the direction of the orbifold action.

The discussion so far applied in most generality only to the Abelian theory with gauge group $U(1)\times U(1)$. In the non-Abelian case the situation is considerably richer: There are still the $1/2$ BPS vortices involving only a single scalar field. Using a gauge transformation it can still be diagonalized and is characterized by $2M$ real numbers ($M\leq N$), the strength of the singularity for this scalar and for the gauge field in $M$ different sub-blocks of $N\times N$ matrices. The M-theory dual is a collection of M2-branes all with the same orientation on $S^7/\bZ_k$ but occupying different $AdS_2\times S^1$ subspaces of $AdS_4$.

The same configuration, where all M2-branes are still oriented the same way, but not along the direction of the Hopf-fiber, is dual to a class of $1/3$ BPS vortices. In the gauge theory description a second scalar field is turned on and is antiholomorphic. The fact that all the M2-branes are aligned is manifested in a constraint on the ratio of the two scalars.

If this constraint is relaxed, we find a more general family of $1/3$ BPS vortices, with $4M$ parameters. Their dual in M-theory is a collection of M2-branes which are not wrapping the same circle on $S^7/\bZ_k$, yet still they are all within an $S^3/\bZ_k$ subspace. It is possible to turn on the two remaining scalars in a way that corresponds to quite general M2-branes on all of $S^7/\bZ_k$, still preserving four supercharges ($1/6$ BPS).

In cases where there is a large number of M2-branes, it is no longer possible to ignore their backreaction and the proper dual description of the vortex loop operator is as a ``bubbling geometry''. The metrics describing the case of the $1/2$ BPS vortex loops are given by orbifolding a known solution \cite{Lunin:2007ab}. A similar analysis should apply also to the $1/3$ BPS ones which have a $1/2$ BPS origin. It would be very interesting to find the more general $1/3$ BPS geometries, those with $4M$ parameters.

Including natural boundary counter-terms for the classical action at weak coupling, 
we got no finite remnants, so the expectation value of the vortex loop operator is 
unity. It should receive   quantum corrections since at strong coupling the 
circular loop operator has the behavior
\be
\vev{V_C}=\exp\left[k\pi\sqrt{\lambda/2}\right].
\ee
It would be interesting to reproduce this from a localization calculation 
similar to that for the BPS Wilson loops 
\cite{Kapustin:2009kz, Drukker:2009hy, Marino:2009jd}.
One can also study other vortex loop operators supported on more complicated 
geometries. We leave this for future exploration.

To get a better handle on these operators we proceeded to calculate their correlation functions with chiral primary local operators. As mentioned above, the similar calculation for the surface operators in $\cN=4$ SYM \cite{Drukker:2008wr} suggested the precise agreement between supergravity and a finite series of quantum corrections to the classical gauge theory results.

In the case of the $\cN=6$ Chern-Simons theory, the results were much more complicated. The correlator has non-trivial dependence on the gauge coupling as well as the parameters of the vortex loop operator which do not agree between weak and strong coupling, meaning that they get renormalized. One feature that can be traced from weak to strong coupling, though, is that in both cases the correlator contains a polynomial of the same degree in the respective couplings.

We would like to point out that this Chern-Simons theory has other loop operators --- Wilson loops. These are order-operators, which can be expressed by the insertion of fundamental fields into the path integral. While the operators presented here have some distinct features that we could compare between the different regimes and they seem quite different from those of the Wilson loops of \cite{Drukker:2008zx,Chen:2008bp,Rey:2008bh}, we cannot be sure that these operators do not mix with each-other.

To conclude, the program of identifying the bulk gravitational description of non-local operators in  ${\cN=6}$   Chern-Simons theory is the three dimensional counterpart of the analogous program for ${\cal N}=4$ SYM. 
There, supersymmetric Wilson loops can be described
in a variety of ways, perturbatively in $\cN=4$ SYM \cite{Erickson:2000af,Drukker:2000rr},
as strings in $AdS_5$  \cite{Rey,Maldacena-wl,Drukker:1999zq,Berenstein:1998ij}, 
as a configuration of D3-branes \cite{Drukker:2005kx,Gomis:2006sb,Gomis:2006im}\ or
as a configuration of D5-branes \cite{Yamaguchi:2006tq,Gomis:2006sb}, 
and finally as  asymptotically $AdS_5\times S^5$ ``bubbling'' supergravity backgrounds 
\cite{Yamaguchi:2006te,Lunin:2006xr,D'Hoker:2007fq}. Likewise disorder surface operators can be
given a probe D3-brane description \cite{Gukov:2006jk,Gomis:2007fi} as a well as a ``bubbling'' supergravity description \cite{Gomis:2007fi}, while order surface operators can be given a probe D7-brane description as well as ``bubbling''
supergravity description \cite{Buchbinder:2007ar,Harvey:2008zz}.

In the context of the $AdS_4$/CFT$_3$ duality,  apart from the dictionary 
proposed already in \cite{Aharony:2008ug}, 
and  the bulk identification of the disorder loop operators found in this paper, the 
D2 and D6 probe brane   description of a family of Wilson loops was found in 
\cite{Drukker:2008zx} (see also \cite{Kluson:2008wn}), while 
the M2-brane giant graviton description of chiral primary operators has appeared in 
\cite{Nishioka:2008ib,Berenstein:2008dc}.  In \cite{Gomis:2008vc} (see also \cite{Terashima:2008sy,Hanaki:2008cu})
the ${\cN=6}$   Chern-Simons theory description of multiple M5-branes was proposed.
These probe branes, and others which may still be found, promise to be useful 
and interesting tools to understand the strong coupling dynamics of 
three-dimensional conformal field theories.

\subsection*{Acknowledgments}
N.D. would like to thank Ofer Aharony, Rajesh
Gopakumar, Oleg Lunin, Juan Maldacena, Shiraz Minwalla, Constantinos
Papageorgakis, Jan Plefka, David Tong, Spenta Wadia, Xi Yin and all the
participants of the Monsoon Workshop for stimulating discussion. 
J.G. would like to thank Shunji Matsuura and Filippo Passerini for 
discussions and collaboration regarding non-local operators in Chern-Simons-matter theories.
N.D. acknowledges the welcome hospitality of the Tata Institute for
Fundamental Research and the ICTS, Mumbai during the course of this
work.  Research at Perimeter Institute for Theoretical Physics is
supported in part by the Government of Canada through NSERC and by the
Province of Ontario through MRI. J.G. also acknowledges
further support from an NSERC Discovery Grant.  D.Y. acknowledges the
support of the National Sciences and Engineering Research Council of
Canada (NSERC) in the form of a Postdoctoral Fellowship, and also
support by the Volkswagen Foundation.

\newpage
\appendix

\section{Superconformal Symmetries}
\label{app-Ss}

In this appendix we show that the conformally invariant vortices that preserve some of the Poincar\'e supersymmetries  also preserve the same amount of conformal supersymmetries. There is a simple proof of this statement using group theory; the superconformal generators are given by the commutator of the special conformal generators and the Poincar\'e supercharges, so are necessarily a symmetry of any operator invariant under the other two generators. Still we find it interesting to go through the exercise in detail, since this theory and its formalism are quite new.

Like in the case of the Poincar\'e supercharges, the only non-trivial superconformal variation in a bosonic background is that of the fermions.  The superconformal transformations are obtained in the usual way once the Poincar\'e supersymmetry variation is known, see \cite{Bandres:2008ry} (we  follow  the convention in    \cite{Terashima:2008sy}). The variation is given by \eqn{conformal}
\begin{eqnarray}
\label{varconf}
\delta \psi_I &=& - \gamma^\mu \gamma^\nu x_\nu\eta_{IJ} D_\mu C^J
+{2 \pi}\gamma^\nu x_\nu \left(
-\eta_{IJ} (C^K C_K^{\dagger} C^J-C^J C_K^{\dagger} C^K)
+2 \eta_{KL} C^K C_I^\dagger C^L\right)
\cr&&
-\eta_{IJ}C^J\,.
\end{eqnarray}
where $\eta^{IJ}$ is a constant spinor satisfying 
 \be
\eta^{IJ}= (\eta_{IJ})^*\,,\qquad
\eta^{IJ} =\frac{1}{2} \eta^{IJKL} \eta_{KL}\,,
\ee
and as with $\epsilon_{IJ}$ \eqn{heli}, we decompose $\eta_{IJ}$ according to their helicity in the $z$-plane, so that  $\eta_{IJ}=\eta^+_{IJ}+\eta^-_{IJ}$, where
\be
\gamma^z\eta^+_{IJ}=0\,,\qquad\qquad
\gamma^{\bar z}\eta^-_{IJ}= 0\,.
\ee

For simplicity we do all the calculations for gauge group $U(1)\times U(1)$ but our analysis will apply for all the solutions discussed in Section~\ref{sec-CS}, since all the matrices there commute. In the Abelian theory (\ref{varconf}) reduce to 
\be
\label{varconfdiag}
\delta \psi_I= - \left(\gamma^\mu \gamma^\nu x_\nu D_\mu C^J+C^J\right)\eta_{IJ}
=- \left(2zD_z C^J+C^J\right)\eta^+_{IJ} 
- \left(2\bar zD_{\bar z} C^J+C^J\right)\eta^-_{IJ}\,.
\ee

For the $1/2$ BPS vortex
\be
 \label{halfcon}
C^1=\frac{\beta}{\sqrt{z}}\,.
\ee
Equation (\ref{varconfdiag}) vanishes then for the following $\eta_{IJ}^\pm$
\be
\left\{\eta^+_{12}\,,
\eta^+_{13}\,,
\eta^+_{14}\,,
\eta^-_{23}\,,
\eta^-_{24}\,,
\eta^-_{34}\right\}.
\label{su-conf-1/2}
\ee

The $1/3$ BPS vortex has
 \be
 \label{thirdcon}
C^1= \frac{\beta_1}{\sqrt{z}}\,,
\qquad 
C^2 =\frac{\beta_2}{\sqrt{\bar{z}}}\,.
\ee
Clearly all the supercharges broken by the $1/2$ BPS vortex are still broken, and there are now similar conditions stemming from $C^2$, with the opposite helicity. Together (\ref{varconfdiag}) vanishes for
\be
\left\{\eta^+_{13}\,,
\eta^+_{14}\,,
\eta^-_{23}\,,
\eta^-_{24}\right\}.
\ee

The analysis for the $1/6$ BPS vortex goes along the same lines, giving two preserved conformal supersymmetries.

\section{Spherical Harmonics and Chiral Primary Operators}
\label{app-harmonics}

In this appendix we study the spherical harmonics on $S^7/\bZ_k$ and in particular 
those invariant under an $SU(3)$ subgroup of the $SU(4)$ symmetry group. These 
spherical harmonics will allow us to construct the chiral primary operators which 
couple to the $1/2$ BPS vortex loop operators $V_C$ and the supergravity 
modes dual to them.

The spherical harmonics of $S^7$ which transform in the $SO(8)$ representation with Dynkin label $[J,0,0,0]$ are homogeneous polynomials of degree $J$ in the complex coordinates (\ref{ws})
\be
\begin{aligned}
w^1&=\sin\frac{\vartheta_1}{2}\,e^{i\xi_1}\,,\qquad
&w^3&=\cos\frac{\vartheta_1}{2}\cos\frac{\vartheta_2}{2}
\sin\frac{\vartheta_3}{2}\,e^{i\xi_3}\,,\\
w^2&=\cos\frac{\vartheta_1}{2}\sin\frac{\vartheta_2}{2}\,e^{i\xi_2}\,,\qquad
&w^4&=\cos\frac{\vartheta_1}{2}\cos\frac{\vartheta_2}{2}
\cos\frac{\vartheta_3}{2}\,e^{i\xi_4}\,,
\label{wsa}
\end{aligned}
\ee
and their complex conjugates. 
These spherical harmonics are   eigenvectors of the $S^7$ Laplacian with   eigenvalue $-J(J+6)$.

Explicitly, we write the spherical harmonics as
\be
Y^A\equiv C^{(A)}{}^{J_1\cdots J_{\Delta_-}}_{I_1\cdots I_{\Delta_+}}\, 
w^{I_1}\cdots w^{I_{\Delta_+}}\bar w_{J_1}\cdots\bar w_{J_{\Delta_-}}
\label{harmon}
\ee
where $J=\Delta_++\Delta_-$ and $C^{(A)}{}^{J_1\cdots J_{\Delta_-}}_{I_1\cdots I_{\Delta_+}}$ is a totally symmetric tensor in $I_1\cdots I_{\Delta_+}$  and 
$J_1\cdots J_{\Delta_-}$ and traceless, {\em i.e.}
\be
C^{(A)}{}^{J_1\cdots J_{\Delta_-}}_{I_1\cdots I_{\Delta_+}}\, \delta^{I_q}_{J_r}=0
\ee
for any $1\leq q\leq\Delta_+$ and any $1\leq r\leq \Delta_-$. 
They are normalized as
\be
C^{(A)}{}^{J_1\cdots J_{\Delta_-}}_{I_1\cdots I_{\Delta_+}}\,
\bar{C}^{(B)}{}^{I_1\cdots I_{\Delta_+}}_{J_1\cdots J_{\Delta_-}}=\delta^{AB}\,.
\label{C-norm}
\ee
$\bZ_k$ acts on all the $w^I$ in \eqn{wsa} by $w^I\to e^{2\pi i/k}w^I$, thus the $S^7$ spherical harmonics which survive the $\bZ_k$ orbifold are those where the difference between the number of holomorphic and anti-homorphic coordinates is an integer multiple of $k$, so $\Delta_+-\Delta-=pk$. We get
\be
\Delta_+=\Delta+\frac{pk}{2}\,,\qquad
\Delta_-=\Delta-\frac{pk}{2}\,,\qquad
\Delta=\frac{J}{2}\,.
\ee

The parametrization \eqn{wsa} makes manifest the embedding $SU(4)\times U(1)_B\subset SO(8)$, where ${\bf 8}_v\rightarrow {\bf 4}_1\oplus {\bf \bar{4}}_{-1}$.
We further consider the decomposition  $SU(3)\times U(1)_R\subset SU(4)$, where ${\bf 4}\rightarrow {\bf 1}_1\oplus {\bf 3}_{-1/3}$ and 
would like to focus now on spherical harmonics invariant under this $SU(3)$ subgroup. 
The $SU(3)$ invariant harmonics, transforming in the $[\Delta_+,0,\Delta_-]$ representation of  $SU(4)$, are functions of $w^1$, $\bar w_1$ and $|w^2|^2+|w^3|^2+|w^4|^2$ only. In terms of the angular coordinates in \eqn{wsa}, we have that the $SU(3)$ invariant spherical harmonics may depend only on $\vartheta_1$ and $\xi_1$

In order to make manifest the $U(1)_B$ and $U(1)_R$ symmetries one may redefine the angles in  \eqn{wsa} as
\be
\xi_1=\frac{\zeta}{k}+\varphi_1\,,\quad
\xi_2=\frac{\zeta}{k}-\frac{\varphi_1}{3}+\varphi_2\,,\quad
\xi_3=\frac{\zeta}{k}-\frac{\varphi_1}{3}-\varphi_2+\varphi_3\,,\quad
\xi_4=\frac{\zeta}{k}-\frac{\varphi_1}{3}-\varphi_3\,.
\ee
The $\zeta$ coordinate parametrizes the Hopf fiber of the $S^7$, so  $U(1)_B$ is generated by $\partial_\zeta$ while 
$U(1)_R$ is generated by $\partial_{\varphi_1}$. The Killing vectors $\partial_{\varphi_2}$ and $\partial_{\varphi_3}$
generate the Cartan subalgebra of $SU(3)$. As mentioned above, the $U(1)_B$ charge of the spherical harmonic is the number of holomorphic coordinates minus the number of antiholomorphic coordinates in the harmonic. The spherical harmonics with zero $U(1)_B$ charge correspond to states that do not carry any angular momentum around the ``M-theory circle'' and remain light in weakly coupled Type IIA string theory.

For practical purposes it is better to continue employing $\vartheta_1$ and $\xi_1$, and write the $S^7$ Laplacian with $SU(3)$ invariance as
\be
\left(\frac{4}{\sin\frac{\vartheta_1}{2}\cos^5\frac{\vartheta_1}{2}}\partial_{\vartheta_1}
\sin\frac{\vartheta_1}{2}\cos^5\frac{\vartheta_1}{2}\,\partial_{\vartheta_1}
+\frac{1}{\sin^2\frac{\vartheta_1}{2}}\partial^2_{\xi_1} \right) Y_{\Delta,p}
=-J(J+6)\,Y_{\Delta,p}\,.
\label{laplace}
\ee
This is solved by
\be
Y_{\Delta,p}(\vartheta_1,\xi_1)={\cal N}_{\Delta,p}\,
\sin^{pk}\frac{\vartheta_1}{2}\,e^{ipk\xi_1}\,
P_{\Delta-\frac{pk}{2}}^{(pk,2)}(\cos\vartheta_1)\,,
\label{sph-k}
\ee
where $\Delta=J/2\geq |pk|/2$. $P_n^{(\alpha,\beta)}$ are Jacobi polynomials, which  we may also write in terms of   hypergeometric functions as
\be
P_{\Delta_-}^{(pk,2)}(\cos\vartheta_1)
=\frac{\Delta_+!}{\Delta_-! (pk)!}\,
{}_2F_1\left(\Delta_++3,\,-\Delta_-\,;1+pk\,;\sin^2\frac{\vartheta_1}{2}\right)\,.
\ee

The normalization constant ${\cal N}_{\Delta,p}$ in \eqn{sph-k} is fixed such that the normalization of $Y^A$ agrees with that which is determined from \eqn{C-norm} to be
\be
\int_{S^7} Y^A\bar{Y}^B
=2\pi^4\,\frac{\Delta_-!\Delta_+!}{(2\Delta+3)!}\,\delta^{AB}
\label{normm}
\ee
where the volume of the unit radius $S^7$ is $\Omega_7=\pi^4/3$ and on $S^7/\bZ_k$ the right-hand side gets a factor of $1/k$.

To  prove this we first use the identity
\be
\int_{S^7} e^{j\cdot\bar w+\bar j\cdot w}
=2\pi^4\sum_{m=0}^\infty\frac{(j\cdot\bar j)^m}{m!(m+3)!}\,.
\ee
Differentiating $m$ times with respect to $j$ and $m$ times with respect to $\bar j$ and setting $|j|=0$, we get
\be
\int_{S^7} w^{I_1}\cdots w^{I_m}\bar w_{J_1}\cdots\bar w_{J_m}
=\frac{2\pi^4}{(m+3)!}\sum_{\sigma\in S_m}
\delta^{I_1}_{J_{\sigma(1)}}\cdots\delta^{I_m}_{J_{\sigma(m)}}\,,
\ee
where the sum is over all permutations. Finally we plug  this formula into the left hand side of  (\ref{normm}), and notice that 
of the $(2\Delta)!$ possible permutations, only $\Delta_-!\Delta_+!$ give a non-zero contraction between the two $C^{(A)}$ tensors, and we get the right-hand side of (\ref{normm}).

The Jacobi polynomials are conventionally normalized as
\be
\int_{S^7}{}\left[\sin^{pk}\frac{\vartheta_1}{2}\,P_{\Delta_-}^{(pk,2)}(\cos\vartheta_1)\right]^2
=\frac{\pi^4}{( 2\Delta+3)}\,
\frac{\Delta_+!(\Delta_-+2)!}{\Delta_-!(\Delta_++2)!}\,.
\ee
Together with equation (\ref{normm}) we find that the $SU(3)$ invariant spherical harmonics that gives rise to unit normalized operators are given by
\be
\label{normalized}
Y_{\Delta,p}(\vartheta_1,\xi_1)
=\sqrt{\frac{2\,(\Delta_++2)!}{(2\Delta+2)!\,(\Delta_-+2)!}}\,(\Delta_-)!\,
\sin^{pk}\frac{\vartheta_1}{2}\,e^{ipk\xi_1}\,
P_{\Delta_-}^{(pk,2)}(\cos\vartheta_1)\,.
\ee
The first few properly normalized harmonics with $p=0$ are given by
\be
\begin{aligned}
Y_{1,0}(\vartheta_1)&={\frac{1}{2\sqrt{3}}}\left(-1+2\cos\vartheta_1\right),\\
Y_{2,0}(\vartheta_1)&={\frac{1}{12\sqrt{10}}}\left(-1-10\cos\vartheta_1+15\cos^2\vartheta_1\right),\\
Y_{3,0}(\vartheta_1)&={\frac{1}{16\sqrt{35}}}\left(3-6\cos\vartheta_1-21\cos^2\vartheta_1+28\cos^3\vartheta\right)\,.
\end{aligned}
\label{pJs}
\ee

These spherical harmonics can be used to write down the $SU(3)$ invariant chiral primary operators. As mentioned in Section~\ref{sec-CS}, the unit normalized chiral primary operators with vanishing $U(1)_B$ charge are given by (\ref{general-local})
\be
{\cal O}^A_{\Delta,0}={(4\pi)^\Delta\over \lambda^{\Delta}\sqrt{\Delta}}
C^{(A)}{}^{J_1\cdots J_{\Delta}}_{I_1\cdots I_{\Delta}}\, 
\hbox{Tr}\Big(C^{I_1}C^\dagger_{J_1}\cdots 
C^{I_{\Delta}}C^\dagger_{J_{\Delta}}\Big),
\label{general-locala}
\ee
Using the embedding coordinates in (\ref{wsa}),  the harmonics in (\ref{pJs}) give the first few unit normalized  $SU(3)\times U(1)_B$ invariant operators%
\footnote{Note that the index $I$ sums over all directions, including 1, and all monomials should be symmetrized.}
\begin{align}
\cO_{1,0}&=\frac{2\pi}{\sqrt{3}\lambda}\Tr\Big[C^IC^\dagger_I-4C^1C_1^\dagger\Big],
\nonumber\\
\label{Os}
\cO_{2,0}&=\frac{8\pi^2}{3\sqrt{5}\lambda^2}
\Tr\Big[(C^IC_I^\dagger)^2-10C^IC_I^\dagger\,C^1C_1^\dagger+15(C^1C_1^\dagger)^2\Big],\\
\cO_{3,0}&=\frac{16\pi^3}{3\sqrt{105}\lambda^3}
\Tr\Big[(C^IC_I^\dagger)^3-18(C^IC_I^\dagger)^2\,(C^1C_1^\dagger)+
63(C^IC_I^\dagger)\,(C^1C_1^\dagger)^2-56(C^1C_1^\dagger)^3\Big].\nonumber
\end{align}

While it is no harder to write down the spherical harmonics with non-zero $U(1)_B$ charge $pk$,  the corresponding gauge invariant local operators are rather subtle objects. The analog of \eqn{general-locala} for non-zero $p$ will have a different number of $C^I$ and $C^\dagger_I$ fields and   cannot be trivially traced over. The rigorous definition of the corresponding operator requires us to include an 't Hooft operator carrying $p$ units of magnetic flux. This object transforms in the $pk$ symmetric product of the bi-fundamental of $U(N)\times U(N)$ and can soak up the color indices on the extra $pk$ fields. Unfortunately, it is not known how to write them down in general.

Still, given that all our classical configurations are made of commuting matrices and that the gauge symmetry is broken --- and being a bit cavalier --- we can try to write down the relevant operators. For example, in the case when $\Delta=pk/2$, using that $P_0^{(\alpha,\beta)}=1$ the properly normalized spherical harmonics are
\be
Y_{\frac{pk}{2},p}(\vartheta_1,\xi_1)=\sin^{pk}\frac{\vartheta_1}{2}e^{ipk\xi_1}\,.
\ee
The operators with $\Delta=pk/2$ are then of the general form
\be
\cO_{\frac{pk}{2}, p}\sim\frac{(4\pi)^{pk/2}}{\lambda^{pk/2}}\, (C^1)^{pk}\,.
\ee

\section{String Theory Description}
\label{app-D2}

For completeness we present here the M2-brane solution of Section~\ref{sec-M2} also in type IIA string 
theory language where it is replaced by a D2-brane. In this case the string background is given by
\be
ds^2_\text{string}=\frac{R^3}{4k}\left(
ds^2_{AdS_4}+4ds^2_{\CP^3}\right).
\ee
For the $AdS_4$ metric we take the same metric as before \eqn{metricmaca}. We describe $\CP^3=S^7/S^1$ by taking the metric \eqn{S7-metric}, isolating the overall phase \eqn{ws}
\be
\zeta=\frac{1}{4}\left(\xi_1+\xi_2+\xi_3+\xi_4\right),
\ee
and defining three other phases as differences of the $\xi_i$. Then the metric on $S^7$ is realized as a Hopf fiber over $\CP^3$
\be
ds^2_{S^7}=ds^2_{\CP^3}+(d\zeta+\omega)^2\,,
\ee
where $d\omega$ is the K\"ahler form on $\CP^3$.

In addition to the metric, the supergravity background has the dilaton, and the two-form and four-form field strengths from the Ramond-Ramond sector
\be
e^{2\Phi}=\frac{R^3}{k^3}\,,
\qquad
F_4=\frac{3}{8}\,R^3\,\Omega_{AdS_4}\,,
\qquad
F_2=k\,d\omega\,.
\ee
Here $\Omega_{AdS_4}$ is the volume form on $AdS_4$.
As in the M-theory description, for the three-form potential we take
\be
C_3=\frac{1}{8}R^3(\cosh^3u-1)\,\Omega_{AdS_2}\wedge d\phi\,.
\ee

This string theory description is valid in the regime
\be
\lambda\gg1\,,\qquad
k^5\gg N\,.
\ee

The M2-brane solutions are contained within an $S^3/\bZ_k\subset S^7/\bZ_k$ and likewise for the D2-branes we take $w^3=w^4=0$ which gives a $\CP^1\subset\CP^3$. 
Parametrizing it by 
\be
w^1=\sin\frac{\vartheta_1}{2}\,e^{-i\frac{\varphi}{2}}\,,\qquad
w^1=\cos\frac{\vartheta_1}{2}\,e^{i\frac{\varphi}{2}}\,,
\ee
gives 
\be
ds_{\CP^1}^2
=\frac{1}{4}\left(d\vartheta_1^2+\sin^2\vartheta_1\,d\varphi^2\right),\qquad
C_1=\frac{k}{2}(\cos\vartheta_1\mp1)d\varphi\,,
\label{C1-gauge}
\ee
where $F_2=dC_1$ and the choice of sign in $C_1$ corresponds to two different gauges 
with the Dirac string at oposite poles. Note that because of the
factor of $1/4$, the radius of $AdS_4$ and of $S^2$ are equal.

Like the M2-brane, the D2-brane will occupy an $AdS_2\times S^1\subset AdS_4$ where we may parameterize $AdS_2$ by either \eqn{poincare-AdS2-metric} or \eqn{global-euclid-metric} and the calculation goes through identically. The $S^1\subset AdS_4$ is parametrized by $\phi$ and we allow the angle $\varphi$ on $\CP^1$ to vary with $\phi$. In principle $u$ and $\vartheta_1$ should be functions on the world-volume, though from symmetry arguments we expect them to be constants.

The action includes the Dirac-Born-Infeld piece and the
Wess-Zumino coupling
\be
\cS_\text{D2}=T_\text{D2}\int e^{-\Phi}\sqrt{\det(g+2\pi\alpha'F)}
-T_\text{D2}\int\Big[P[C_3]+2\pi i\alpha'P[C_1]\wedge F\Big].
\label{D2-action}
\ee
Here $g$ is the induced metric on the world-volume and $F$ is the
gauge field. The vortex may carry electric flux, which by symmetry is proportional to 
the volume form on $AdS_2$, $F=E\, \Omega_{AdS_2}$. Being an electric field 
in a theory with Euclidean signature, $E$ is imaginary. $P[C_3]$
is the pullback of the Ramond-Ramond three-form potential and $P[C_1]$ that of
the one-form. The last term comes with an $i$ again due to the fact that we are in Euclidean signature.

Plugging our ansatz in we find
\be
\begin{aligned}
\cS_\text{D2}=\frac{T_\text{D2}R^3}{8}\int \Omega_{AdS_2}\, d\phi
\bigg[&\sqrt{(\cosh^4u+\tau^2E^2)(\sinh^2u+\dot\varphi^2\sin^2\vartheta_1)}
\\&
-\cosh^3u+1-i\dot\varphi\,\tau E(\cos\vartheta_1-1)\bigg]\,,
\end{aligned}
\ee
with $\tau=8\pi k/R^3=\sqrt{2/\lambda}$ (setting $\alpha'=1$) and in 
our conventions $T_\text{D2}=1/(4\pi^2)$.

The equation of motion for $u$ leads to the two possible values of $E$
\begin{align}
&1.\qquad
i\tau E=\cosh u\,\sqrt{1-\dot\varphi^2\sin^2\vartheta_1}\,,
\label{D2-sol-1}
\\
&2.\qquad
i\tau E=\cosh u\sqrt{4(1-\dot\varphi^2\sin^2\vartheta_1)-3\cosh^2 u}\,.
\label{D2-sol-2}
\end{align}
Only the first of these two solutions seems to be related to the vortex loop operators and is the analog of \eqn{u-eom}.

Concentrating on \eqn{D2-sol-1}, the $\vartheta_1$ equation of motion again has two solutions. The first one has $\dot\varphi=0$, in complete analogy with \eqn{M2-1/2}. This solution preserves 12 supercharges and is the string theory dual of the $1/2$ BPS vortex loop.

The other solution has
\be
\dot\varphi=\pm1\,,\qquad
i\tau E=\mp\cosh u\cos\vartheta_1\,.
\ee
This is the analog of the M2-brane solution \eqn{M2-1/3} and preserves eight supercharges.

Note that for both the $1/2$ BPS and $1/3$ BPS solutions the values of $u$ and of $\vartheta_1$ are free parameters, not constrained by the equations of motion.

The gauge field is a cyclic variable and the flux through the brane is
proportional to the conjugate momentum
\be
p=-2\pi i \,\frac{\delta\cL}{\delta F}
=\pm 2\pi^2kT_{D2}=\pm\frac{k}{2}\,.
\label{p-D2}
\ee
This flux should be integer quantized, which happens only for even $k$. This is the string theory manifestation of the fact that a single vortex loop operator is not well defined for odd $k$.%
\footnote{Note also that due to the existence of 't Hooft operators, the electric flux is 
defined only modulo $k$, which is manifested here in the two gauge choices for $C_1$ \eqn{C1-gauge}.}

To summarize, the most general D2-brane solution has the following parameters: $u$, $\vartheta_1$, $\varphi_0$, where $\varphi=\varphi_0\pm\phi$ and since the world-volume has a compact direction we can have a holonomy for the $U(1)$ gauge field around it $A_\phi$. They are related to the parameters of the $1/3$ BPS vortex loop operator by \eqn{map}
\be
\sinh^2u=\frac{|\beta_1|^2+|\beta_2|^2}{2\pi^2\lambda}\,,
\qquad
\tan\frac{\vartheta_1}{2}\,e^{-i\varphi_0}=\frac{\beta_1}{\beta_2}\,,
\qquad
A_\phi=\alpha\,.
\ee

Finally we evaluate the action on this classical solution. 
As is explained in \cite{Drukker:2005kx}, the action as it stands will
not give the correct classical value, since it is a functional of the
electric field. The action should be a functional of the conserved
quantity which gives a good variational problem. This is the flux
conjugate to the gauge field, namely $p$. We therefore have to perform
a Legendre transform
\be
\cS_\text{L.T.}=\cS-i\int \frac{p}{2\pi}\, F\,.
\ee
For the solution of interest \eqn{D2-sol-1}, the action is proportional to the volume of 
$AdS_2$. In the case of the circular loop operator the regularized area is 
$-2\pi$ and we find
\be
\cS_\text{D2}^\text{classical}=\frac{T_\text{D2}R^3}{8}\int \Omega_{AdS_2}\, d\phi
=-\frac{R^3}{8}=k\pi\sqrt{\lambda/2}\,,
\ee
Exactly as in the M-theory calculation \eqn{final-M}.

\section{Supersymmetry of Brane Solution}
\label{app-M2-SUSY}

In this appendix we show that the M2-brane solutions presented in Section~\ref{sec-M} indeed preserve $1/2$ and $1/3$ of the supercharges.

\subsection{Killing Spinors}
\label{app-killing}

To check the supersymmetries preserved by the brane solution we need an 
explicit form of the Killing spinors on $AdS_4\times S^7/\bZ_k$. For the $AdS_4$ part we take \eqn{metricmaca} but with the $AdS_2$ factor being global Lorentzian $AdS_2$ 
\be
ds_{AdS_4}^2=du^2+\cosh^2u\left(d\rho^2-\cosh^2\rho\,dt^2\right)+\sinh^2u\,d\phi^2\,,
\ee
For $S^7$ we take \eqn{S7-metric}.

We choose the elfbeine to be
\be
\begin{gathered}
e^0 = \frac{R}{2} \cosh u \cosh \r \, dt\,, \quad 
e^1 = \frac{R}{2} \cosh u \, d\r\,, \quad
e^2 = \frac{R}{2} du \,, \quad 
e^3 = \frac{R}{2} \sinh u \, d\phi\,,
\\
e^4 =\frac{R}{2}  d\vartheta_1\,, \qquad 
e^5 =\frac{R}{2}  \cos\frac{\vartheta_1}{2}\,d\vartheta_2\,,\qquad 
e^6 =\frac{R}{2}  \cos\frac{\vartheta_1}{2}\cos\frac{\vartheta_2}{2} \, d\vartheta_3\,,
\\
e^7=R\sin\frac{\vartheta_1}{2}\,d\xi_1\,,\qquad
e^8=R\cos\frac{\vartheta_1}{2}\sin\frac{\vartheta_2}{2}\,d\xi_2\,,
\\
e^9=R\cos\frac{\vartheta_1}{2}\cos\frac{\vartheta_2}{2}\sin\frac{\vartheta_3}{2}\,d\xi_3\,,\qquad
e^\natural=R\cos\frac{\vartheta_1}{2}\cos\frac{\vartheta_2}{2}\cos\frac{\vartheta_3}{2}\,d\xi_4\,.
\end{gathered}
\label{beine}
\ee
The Killing spinor equation in this background can be written as
\be
D_M \epsilon = \frac{1}{2} \hat \gamma \gamma_M \epsilon
\ee
where the index $M$ runs over all 11 coordinates, and $\hat \gamma =\gamma^{0123}$. Note that 
small $\gamma$ have tangent-space indices while capital $\Gamma$ carry 
curved-space indices.

The Killing spinors that solve this equation are \cite{Nishioka:2008ib,Drukker:2008zx}
\be
e^{\frac{\vartheta_1}{4}\hat \gamma \gamma_4}
e^{\frac{\vartheta_2}{4}\hat \gamma \gamma_5}
e^{\frac{\vartheta_3}{4}\hat \gamma \gamma_6}
e^{\frac{1}{2}(\xi_1\gamma_{47}+\xi_2\gamma_{58}
+\xi_3\gamma_{69}+\xi_4\hat \gamma \gamma_\natural)}
e^{\frac{u}{2} \hat \gamma \gamma_2}
e^{\frac{\r}{2} \hat \gamma \gamma_1}
e^{\frac{t}{2} \hat \gamma \gamma_0}
e^{\frac{\phi}{2} \gamma_{23}}\epsilon_0
={\mathcal M}\epsilon_0
\label{Killing}
\ee
$\epsilon_0$ is a constant 32-component spinor 
and the Dirac matrices were chosen such that
$\gamma_{012345678 9\natural}=1$. 
A similar calculation in a different coordinate system was done in 
\cite{Nishioka:2008ib}.

Recall that the angles $\xi_i$ have period $2\pi$ up to the $\bZ_k$ orbifold, which acts on all by $\xi_i\to\xi_i+2\pi/k$. We have to check whether the Killing spinors are invariant under this action and survive the orbifold projection. To do this it is convenient to write the spinor $\epsilon_0$ in a basis which diagonalizes
\be
i\gamma_{47}\epsilon_0=s_1\epsilon_0\,,\qquad
i\gamma_{58}\epsilon_0=s_2\epsilon_0\,,\qquad
i\gamma_{69}\epsilon_0=s_3\epsilon_0\,,\qquad
i\hat\gamma\gamma_\natural\epsilon_0=s_4\epsilon_0\,.
\label{ss}
\ee
All the $s_i$ take values $\pm1$ and by our conventions on the product 
of all the Dirac matrices, the number of negative eigenvalues is even. 
Now consider the orbifold action, the Killing spinors transform as
\be
{\mathcal M}\epsilon_0\to {\mathcal M} e^{i\frac{\pi}{k}
(s_1+s_2+s_3+s_4)}\epsilon_0\,.
\ee
This transformation is a symmetry of the Killing spinor when two of the $s_i$ eigenvalues are positive and two negative and not when they all have the same sign (unless $k=1$ or $k=2$). The allowed values of the $s_i$ are therefore
\be
(s_1,s_2,s_3,s_4)\in\left\{
\begin{matrix}
(+,+,-,-),\ (+,-,+,-),\ (+,-,-,+),\\
(-,+,+,-),\ (-,+,-,+),\ (-,-,+,+)
\end{matrix}
\right\}.
\label{signs}
\ee
Each configuration represents four supercharges, so the orbifolding 
breaks $1/4$ of the supercharges (except for $k=1,2$) and leaves 
24 unbroken supersymmetries.

\subsection{Projector Equation}

The supersymmetry projector equation associated with an M2-brane with world-volume coordinates $t$, $\rho$ and $\phi$ is given by
\be
\frac{1}{\cL_{NG}} \, \partial_t X^M \,\partial_\rho X^N \, \partial_\phi X^L \, 
\Gamma_{MNL} \, \epsilon = \epsilon\,,
\ee
where $M,N,L$ are target-space coordinates and ${\cal L}_{NG}$ is the Langrangian of the membrane, without the Wess-Zumino term.

The M2-brane ansatz involved motion on a subspace of $S^7/\bZ_k$, which for convenience we take here to be that with $\vartheta_1=\vartheta_2=0$ (instead of $\vartheta_2=\pi$ as in Section~\ref{sec-M2}). The remaining coordinates can be defined as $\zeta=\frac{k}{2}(\xi_3+\xi_4)$ and $\varphi=\xi_3-\xi_4$, which were both functions of $\phi$, and $\vartheta=\vartheta_3$ is a constant. The projector equation becomes
\be
\gamma_{01} \left(\sinh u \, \gamma_3 
+\gamma_{\natural}\left(\frac{2}{k}\,\dot\zeta \,e^{-\frac{\vartheta}{2}\gamma_{9\natural}}
-\dot\varphi\, e^{\frac{\vartheta}{2}\gamma_{9\natural}}\right)
\right)\epsilon = \cosh u \, \epsilon\,.
\label{projector}
\ee

Using the relations
\be
\begin{aligned}
&&{\cal M}^{-1} \, \gamma_{01\natural}\,e^{\pm\frac{\vartheta}{2}\gamma_{9\natural}} \,\cM
&= AB^{-1}\,e^{-\frac{\vartheta}{2}(\hat\gamma\gamma_6\pm\gamma_{9\natural})}\,B\,
\gamma_{01\natural}\,,
\\
&&A\equiv \cM^{-1}\,e^{-u\hat\gamma\gamma_2}\,\cM&=
\cosh u-\sinh u\,{\cal M}^{-1} \, \gamma_{013} \, {\cal M}\,,
\\
&&B\equiv e^{\frac{1}{2}(\xi_3 \gamma_{69}+\xi_4\hat \gamma \gamma_\natural)}\,,
\end{aligned}
\ee
the projector equation multiplied from the left by $\cM^{-1}$ can be repackaged as
\be
A\left(
1-B^{-1}\left(
\frac{2}{k}\,\dot\zeta\,e^{-\frac{\vartheta}{2}(\hat\gamma\gamma_6-\gamma_{9\natural})}
-\dot\varphi\,e^{-\frac{\vartheta}{2}(\hat\gamma\gamma_6+\gamma_{9\natural})}\,
\right)B\,\gamma_{01\natural}\right)
\epsilon_0=0\,.
\label{proj-2}
\ee

In the case when $\vartheta=0$ and $\dot\varphi=0$, this reduces to
\be
A\left(1-\frac{2}{k}\,\dot\zeta\,\gamma_{01\natural}\right)
\epsilon_0=0\,.
\label{proj-1/2}
\ee
This has solutions when $\dot\zeta=\pm k/2$, which indeed is the classical solution \eqn{M2-1/2}. This is a single condition on $\epsilon_0$. Furthermore, note that the projector equation \eqn{proj-1/2} commutes with the orbifolding condition \eqn{ss}, \eqn{signs} so for $k=1,2$ there are 16 preserved supercharges, while for general $k$ there are 12. In all cases this is $1/2$ BPS

The second solution \eqn{M2-1/3} has $\dot\varphi=1$ and a constant $\zeta$, which for simplicity we take to be $\zeta=0$. In that case \eqn{proj-2} gives
\be
A\left(1+e^{-\frac{\varphi}{4}(\gamma_{69}-\hat \gamma \gamma_\natural)}\,
e^{-\frac{\vartheta}{2}(\hat\gamma\gamma_6+\gamma_{9\natural})}\,
e^{\frac{\varphi}{4}(\gamma_{69}-\hat \gamma \gamma_\natural)}\,\gamma_{01\natural}\right)
\epsilon_0=0\,,
\ee
which can be rewritten as
\be
\frac{1}{2}\,A\left(2
+(\gamma_{01\natural}-\gamma_{2369})
+\left(\cos\vartheta
+\sin\vartheta\,\gamma_{9\natural}\,e^{\frac{\varphi}{2}(\gamma_{69}-\hat \gamma \gamma_\natural)}\,
\right)(\gamma_{01\natural}+\gamma_{2369})\right)
\epsilon_0=0\,.
\label{proj-1/3}
\ee
One way of solving this equation is by imposing the two conditions
\be
\gamma_{2369}\,\epsilon_0=
-\gamma_{01\natural}\,\epsilon_0=
\epsilon_0\,.
\label{1/3BPS}
\ee
Note that as before we have to take a specific eigenvalue for $\gamma_{01\natural}$ (here with the opposite sign) and now also for $\gamma_{2369}$, which relates the motion along the $\varphi$ circle with $\phi$. The two conditions together give
\be
\gamma_{69}\,\epsilon_0=-\hat\gamma\gamma_\natural\,\epsilon_0\,.
\ee
This is represented in the basis \eqn{ss} as $s_3=-s_4$. Of the six possible combinations of signs in \eqn{signs}, four are allowed
\be
(s_1,s_2,s_3,s_4)\in
\Big\{ (+,-,+,-)\,,(+,-,-,+)\,,(-,+,+,-)\,,(-,+,-,+)\Big\}.
\ee
Each of the sign combinations represents four supercharges, but the extra condition on $\gamma_{01\natural}$ in \eqn{1/3BPS}, reduces the counting by a half. Therefore this M2-brane solution preserves eight supercharges, i.e. it is $1/3$ BPS.

Let us look for other solutions to \eqn{proj-1/3}, where we impose the complementary condition
\be
\gamma_{69}\,\epsilon_0=\hat\gamma\gamma_\natural\,\epsilon_0\,.
\ee
Equation \eqn{proj-1/3} now becomes
\be
\left(1+e^{\vartheta \gamma_{9\natural}}\,\gamma_{01\natural}\right)
\epsilon_0=0\,.
\label{proj-1/6}
\ee
These two equations commute, so it would seem that this brane solution has more than eight preserved supercharges. Note however that unless $\vartheta=0$, equation \eqn{proj-1/6} does not commute with $\gamma_{69}$ and $\hat\gamma\gamma_\natural$, so the solutions will mix the states with eigenvlues
\be
(s_1,s_2,s_3,s_4)\in
\Big\{ (+,+,+,+)\,,(+,+,-,-)\Big\}.
\ee
and likewise the two possibilities with $s_1=s_2=-1$. Therefore equation \eqn{proj-1/6} has no solutions (for $\vartheta\neq0$) on the subspace of states \eqn{signs} preserved by the orbifold, for $k>2$. For $k=1$ and $k=2$ the states with all positive or all negative $s_i$ are allowed and there are eight more solutions to the projector equation. Together with the above there will be a total of 16 supercharges, so for $k=1,2$ it is $1/2$ BPS, just like the solution with $\vartheta=0$.

\section{Supergravity Modes on $AdS_4\times S^7/\bZ_k$}
\label{app-SUGRA-modes}

In this appendix we present the necessary ingredients of the 
fluctuation spectrum of  eleven dimensional supergravity around the 
$AdS_4\times S^7/\bZ_k$ vacuum that are needed for the calculation of 
the correlation functions of the vortex loop operators and chiral 
primary operators in the probe approximation in supergravity, 
as performed  in Section~\ref{sec-local-M}.

The required formalism of the fluctuations around the $AdS_4 \times S^7$ 
supergravity background were developed in 
\cite{Biran:1983iy,Castellani:1984vv,Bastianelli:1999bm} of which 
we follow mainly \cite{Bastianelli:1999bm} with some necessary modifications. 

Using late greek letters $\m,\n,\cdots$ for the $AdS_4$ portion of the metric 
and early greek letters $\a,\b,\cdots$ for the $S^7$ we expand the metric 
$\tilde g$ and three-form $\tilde C$ about the $AdS_4\times S^7$ 
background $g$ and $C$ in terms of the 
fluctuations modes $h_{\mu\nu}$, $H_{\mu\nu}$, $h_{\alpha\beta}$, $\pi$, 
$\delta C_{\mu\nu\rho}$ and $b$ as
\be
\begin{gathered}
\label{SUGfluc}
\tilde g_{\m\n} = g_{\m\n} + h_{\m\n}\,,\qquad 
\tilde g_{\a\b} =g_{\a\b} + h_{\a\b}\,,\\ 
h_{\m\n} = H_{\m\n} - \frac{1}{2} g_{\m\n}\pi\,,\qquad 
\pi \equiv g^{\a\b} h_{\a\b}\,,\qquad 
H^\m_\m =\frac{9}{7}\pi\,,\\ 
\tilde C_{\m\n\r} = C_{\m\n\r} + \d C_{\m\n\r}
\simeq C_{\m\n\r} - \varepsilon_{\m\n\r\l} \nabla^\l b\,.
\end{gathered}
\ee
The fluctuations of the three-form field $C_{\m\n\r}$ were not provided in
\cite{Bastianelli:1999bm}. Rather, the field $b$ was used to parameterize the 
fluctuations of the dual six-form. Below, in Appendix~\ref{app-C3f} we 
derive the expression for the fluctuation of the three-form field given above 
by application of the constraint relating the three-form and six-form fields
of 11-dimensional supergravity and using the approximation \eqn{SUGmodes}.

The fields are expanded in a Kaluza-Klein
expansion on the $S^7$, giving for example
\be\label{SUGfields}
\pi(x,y) = \sum_A \pi^A(x) Y^A(y), \qquad b(x,y) = \sum_A b^A(x) Y^A(y) 
\ee 
where $x$ are coordinates on $AdS_4$ and $y$ are those on an $S^7$ of
radius 2, so now the equations like \eqn{laplace} are rescaled by $1/4$
\be
\nabla^\a \nabla_\a Y^A = -\frac{1}{4} J(J+6) Y^A\,,
\ee
We consider only the modes that survive the $\bZ_k$ projection and whose properties we studied in Appendix~\ref{app-harmonics}. They are labeled by two quantum numbers $\Delta_\pm$ or $(J,p)$ such that $J=\Delta_++\Delta_-=2\Delta$ and $\Delta_+-\Delta_-=pk$. Accounting for the radius of the sphere and the orbifold projection, they are normalized by \eqn{normm}
\be
\int_{S^7/\bZ_k} Y^A\bar{Y}^B
=\frac{2^8\pi^4}{k}\,\frac{\Delta_-!\Delta_+!}{(2\Delta+3)!}\,\delta^{AB}\,.
\ee
The equations of motion for the $\pi^A$ and $b^A$ fields on $AdS_4$ are mixed
and can be diagonalized into two mass
eigenstates, of which we concern ourselves only with the
lighter%
\footnote{Note that we scaled the form-fields by $1/\sqrt{2}$ 
compared to \cite{Bastianelli:1999bm} in order to be consistent
with the standard Wess-Zumino coupling of the M2 brane used here.} 
one $s^A(x)$ with $J\geq2$ and mass $m_{S^A}^2=J(J-6)/4$. 
Ignoring the contribution from the heavier field we may write 
the modes $\pi^A$, $b^A$ and $H_{\mu\nu}^A$ in terms of 
$s^A$ as
\be \label{SUGmodes}
\begin{gathered}
\pi^A(x)\simeq \frac{7J}{3} s^A(x)\,, \qquad 
b^A(x) \simeq -2 s^A(x)\,, \\
H^A_{\m\n}(x)\simeq \frac{4}{(J+2)}\left[ \nabla_\m \nabla_\n
  +\frac{J(J+6)}{8} g_{\m\n} \right] s^A(x).
\end{gathered}
\ee
Finally we note that as in equation (20) of \cite{Castellani:1984vv}, the
non-trace piece of the $S^7$ metric fluctuations are heavier than 
$s^A(x)$ and so we take
\be
h_{\a\b} \simeq \frac{1}{7}\, g_{\a\b}\, \pi(x).
\label{halphabeta}
\ee

The quadratic action for the $s^A(x)$ field is given
by \cite{Bastianelli:1999bm}
\be
\begin{aligned}
S_{\text{quad.}} = \frac{1}{4\k^2}&
\sum_A \frac{2^8\pi^4}{k}\,\frac{\Delta_-!\Delta_+!}{(2\Delta+3)!}
\frac{2(J+3)J(J-1)}{(J+2)} \\
&\times \int_{AdS_4} d^4 x \sqrt{\det g_{\m\n}}
\left[ - \frac{1}{2} \nabla^\m s^A \nabla_\m s^A - \frac{1}{2}
    m_{s^A}^2 s^A s^A \right],
\end{aligned}
\ee
where in units where $l_p=1$
\be
\frac{1}{4\k^2} = \frac{1}{(2\pi)^8}\left(\frac{R}{2}\right)^9\,.
\ee
From this the bulk-to-bulk propagator may be derived (see for
example \cite{Berenstein:1998ij}) 
\bsp\label{b2b}
\vev{s^A(x)\, s^B(x')} = &\,\frac{\delta^{JB}\G(\D)}{2\pi^{3/2} \G(\D-1/2)}
\frac{k\,\k^2(2\Delta+2)!\, (\Delta+1)}{2^7\pi^4\Delta_-!\Delta_+!\,\Delta\,(2\Delta-1)} \\
&\qquad\times W^\D\, {}_2 F_1(\D,\D-1\,;2\Delta-2\,;-4W)
\end{split}
\ee
where $W$ is the geodesic distance between the two points. For $AdS_4$ parameterized by $ds^2 = (dy^2 + d{\vec x}^2)/y^2$, it is given by
\be
W = \frac{y y'}{(y-y')^2 + (\vec x - \vec x')^2}.
\ee
The bulk-to-boundary propagator is then obtained in the usual way by taking $y\to0$ while scaling the propagator by $1/y^\Delta$. The
correct normalization corresponding to unit normalized operators in the dual 
conformal field theory is
the square-root of that for the bulk-to-bulk propagator \cite{Berenstein:1998ij}. We therefore
have that the bulk-to-boundary propagator is given by
\be
G = c_J \, \frac{y'^\D }{\big((y-y')^2 + (\vec x - \vec x')^2\big)^\D},
\label{Poincare-prop}
\ee
where
\be\label{cJ}
c_J^2 = 
\frac{k\,\kappa^2}{2^8\pi^{11/2}}
\frac{(\Delta-1)!\,(2\Delta+2)!\,(\Delta+1)}
{\Gamma(\Delta-1/2)\,\Delta_-!\,\Delta_+!\,\Delta\,(2\Delta-1)}
=\frac{2^{2\Delta+7}\pi^2k}{R^9}
\frac{(\Delta+1)!^2\,(2\Delta+1)}{\Delta^2\,\Delta_-!\,\Delta_+!}\,.
\ee

To write the propagator in the coordinate system 
\eqn{metricmaca}, \eqn{poincare-AdS2-metric}, we use polar coordinates $ds^2=dt^2+dr^2+r^2\,d\phi^2$ on $\bR^3$ and substitute in equation 
\eqn{Poincare-prop} $y=z/\cosh u$ and $r=z\tanh u$. This gives the propagator 
\eqn{prop} used in Section~\ref{sec-local-M}.

\subsection{Three-Form Fluctuation}
\label{app-C3f}

In \cite{Bastianelli:1999bm} the fluctuations of the three-form field ${C_3}_{\mu\nu\rho}$ 
which are required for our current analysis were not studied. Instead the 
fluctuations of the dual six-form were represented in terms of a field $b$
\be
\delta C_6 = \varepsilon_{\alpha_1 \cdots \alpha_6\beta}\,  \nabla^\beta b\,.
\ee
We derive here the third line of \eqn{SUGfluc}, by using the constraint 
relating $C_6$ and $C_3$ (see \cite{Chen:2007zzr} for a similar
calculation in the context of $AdS_7 \times S^4$)
\be\label{SUGcons}
F_4 + \star H_7 = 0\,,\qquad 
F_4 \equiv d C_3\,,\qquad 
H_7 \equiv dC_6 + \frac{1}{2}C_3 \wedge F_4\,,
\ee
where $\star$ indicates the Hodge dual. The $H_7$ field is proportional to the volume 
form on $S^7$
\be
H_7 = 3\, \varepsilon_{\alpha_1\cdots \alpha_7}
\ee

The fluctuations of $H_7$ can be written as
\be
\label{deltaH}
\d H_7 = d(\d C_6)
= \varepsilon_{\alpha_1 \cdots \alpha_6\beta}\,\nabla^\beta \nabla_{\mu} b
+\varepsilon_{\alpha_1 \cdots \alpha_7}\nabla^\beta \nabla_\beta b\,. 
\ee
The fluctuations of $F_4$ are then given by \eqn{SUGcons}
\be
\delta F_4=-\delta(\star H_7)\,.
\ee
This will include the Hodge dual of $\delta H_7$ \eqn{deltaH} and in addition also 
the variation of the measure factor in the Hodge duality acting on $H_7$. 
Since $H_7$ has all its indices in the $S^7$ directions, and its dual 
has all $AdS_4$ directions, the epsilon tensor relating the two 
scales like $\sqrt{\det(g_{\mu\nu})/\det(g_{\alpha\beta})}$. Its variation is
\be
\delta \varepsilon^{\alpha_1\cdots\alpha_7}{}_{\mu_1\cdots\mu_4}
=\frac{1}{2}\left(h^\mu_\mu - h^\alpha_\alpha\right)
\varepsilon^{\alpha_1\cdots\alpha_7}{}_{\mu_1\cdots\mu_4}
=-\frac{6}{7}\pi\,\varepsilon^{\alpha_1\cdots\alpha_7}{}_{\mu_1\cdots\mu_4}
\ee
Together we find (note that the Hodge dual changes the sign of the second term)
\be
\delta F_4=\left(\frac{18}{7}\,\pi-\nabla^\beta \nabla_{\beta} b\right) 
\varepsilon_{\mu_1\cdots\mu_4}
+\varepsilon_{\mu_1 \mu_2\mu_3\nu}\,\nabla^\nu \nabla_\alpha b\,.
\label{deltaF4}
\ee

In the approximation which identifies $b$ with $-2s$ \eqn{SUGmodes}, the 
term in parenthesis in \eqn{deltaF4} can be expressed as
\be
\frac{18}{7}\,\pi-\nabla^\beta \nabla_{\beta} b
\simeq
\nabla^\nu\nabla_\nu b\,.
\ee
Now we can integrate $\delta F_4$ to find
\be
\delta C_{\mu_1\mu_2\mu_3}
\simeq -\varepsilon_{\mu_1 \mu_2\mu_3\nu}\,\nabla^\nu b\,.
\ee


\providecommand{\href}[2]{#2}\begingroup\raggedright\endgroup

\end{document}